\documentclass[a4paper,english,jou]{paper}
\usepackage[T1]{fontenc}
\usepackage[latin9]{inputenc}
\usepackage{color}
\definecolor{shadecolor}{rgb}{0.800781, 0.800781, 0.800781}
\usepackage{array}
\usepackage{units}
\usepackage{framed}
\usepackage{amsthm}
\usepackage{amsmath}
\usepackage{graphicx}

\makeatletter

\providecommand{\tabularnewline}{\\}

\numberwithin{equation}{section}
\numberwithin{figure}{section}

\usepackage{multirow}

\makeatother

\usepackage{babel}
\begin{document}

\title{Elements of Game Theory}

\subtitle{Part I: Foundations, acts and mechanisms.}

\author{Harris V. Georgiou (MSc, PhD)\thanks{Email: harris@xgeorgio.info --- URL: http://xgeorgio.info}}

\institution{\emph{Department of Informatics and Telecommunications,}\\
\emph{National \& Kapodistrian University of Athens, Greece.}}
\maketitle
\begin{abstract}
In this paper, a gentle introduction to Game Theory is presented in
the form of basic concepts and examples. Minimax and Nash's theorem
are introduced as the formal definitions for optimal strategies and
equilibria in zero-sum and nonzero-sum games. Several elements of
cooperative gaming, coalitions, voting ensembles, voting power and
collective efficiency are described in brief. Analytical (matrix)
and extended (tree-graph) forms of game representation is illustrated
as the basic tools for identifying optimal strategies and ``solutions''
in games of any kind. Next, a typology of four standard nonzero-sum
games is investigated, analyzing the Nash equilibria and the optimal
strategies in each case. Signaling, stance and third-party intermediates
are described as very important properties when analyzing strategic
moves, while credibility and reputation is described as crucial factors
when signaling promises or threats. Utility is introduced as a generalization
of typical cost/gain functions and it is used to explain the incentives
of irrational players under the scope of ``rational irrationality''.
Finally, a brief reference is presented for several other more advanced
concepts of gaming, including emergence of cooperation, evolutionary
stable strategies, two-level games, metagames, hypergames and the
Harsanyi transformation.\end{abstract}
\begin{keywords}
Game Theory, Minimax theorem, Nash equilibrium, coalitional gaming,
indices of power, voting ensembles, signaling, bluff, credibility,
promises, threats, utility function, two-level games, hypergames,
evolutionary stable strategies, Harsanyi transformation, metagames.

\vspace{12mm}

\end{keywords}
\noindent {\LARGE{}G}AME THEORY is a vast scientific and research
area, based almost entirely on Mathematics and some experimental methods,
with applications that vary from simple board games to Evolutionary
Psychology and Sociology-Biology in group behavior of humans and animals.
Conflict situations are presented everywhere in the real world, every
day, for thousands of years - not only in human societies but also
in animals. The seller and the buyer have to come up with a mutually
acceptable price for the grocery. The employer and the employee have
to bargain in order to reach a mutually satisfying value for the salary.
A buyer in an auction has to continuously estimate the cost/gain value
of making (or not) the next higher bid for some object. The primary
adversaries in a wolf pack have to decide when it is beneficial to
fight over the leadership and when to stop before they are severely
wounded. A swarm of fish has to collectively ``decide'' what is
the optimal number and distance of the piket members or ``scouts''
that serve as the early warning for the group, perhaps even self-sacrificing
if required. All these cases are typical examples, simpler or more
complex, of conflict situations that depend on bargaining, coordination
and evolutionary optimization. Game Theory provides a unified framework
with robust mathematical foundations for the proper formulation and
analysis of such systems.

\section{The building blocks}

In principle, the mathematical theory of games and gaming was first
developed as a model for situations of conflict. \emph{Game Theory}
is the area of research that provides mathematical formulations and
a proper framework for studying adversarial situations. Although E.
Borel looked at similar problems in the 1920s, John Von Neumann and
Oskar Morgenstern provided two breakthrough papers (1928, 1937) as
a kick-start of the field. Since the early 1940\textquoteright s,
with the end of World War II and stepping into the era of the Cold
War that followed, the work of Von Neumann and Morgenstern has provided
a solid foundation for the most simple types of games, as well as
analytical forms for their solutions, with many applications to Economics,
Operations Research and Logistics. However, there are several limitations
that fail to explain various aspects of real-world conflicts \cite{Luce1957},
especially when the human factor is a major factor. The application
of game-theoretic formulations in designing experiments in Psychology
and Sociology is usually referred to as \emph{gaming} \cite{Thomas1984,Camerer2002}.

\subsection{Games, strategies and solutions}

The term \emph{game} is the mathematical formulation of adversarial
situations, where two or more \emph{players} are involved in competitive
or cooperative acts. The \emph{zero-sum} games are able to model situations
of conflict between two or more players, where one\textquoteright s
gain is the other\textquoteright s loss and vice versa. Most military
problems can be modeled as some form of two-player zero-sum game.
When the structure of the game and the rationale of the players is
known to all, then the game is one of \emph{complete information},
while if some of these information is somehow hidden or unknown to
some players, it is one of \emph{incomplete information}. Furthermore,
if all players are fully informed about their opponents\textquoteright{}
decisions, the game is one of \emph{perfect information}. In contrast,
if some of the information about the other players' moves, the game
is one of \emph{partial }or\emph{ imperfect information}. Such games
of both complete and perfect information are all board games, like
Chess, Go and Checkers, and they are all zero-sum by nature. 

Von Neumann and Morgenstern \cite{JvonNeumann1947} proved that there
is at least one optimal plan of decisions or \emph{strategy} for each
player in all zero-sum games, as well as a \emph{solution} to the
game that comes naturally as a result of all players following their
optimal strategies. At the game\textquoteright s solution, each player
can guarantee that the maximum gain an opponent can gain is kept under
a specific minimal limit, defined only by this player\textquoteright s
own strategy. This assertion was formulated as a theorem called \emph{Minimax}
and in the simple case of two opposing players with only two strategies
each the Minimax solution of the game can be calculated analytically
as a solution of a 2x2 set of linear equations, which determine the
stable solution or \emph{saddle-point}.

The consequences of the Minimax theorem have been thoroughly studied
for many years after its proof. As an example, it mathematically proves
the assertion that all board games, including the most complex ones
like Chess, have at least one solution, i.e., an optimal (pure) strategy
for both players that can be analytically calculated, at least in
theory \cite{Stahl1999,Thomas1984,Prywes1999}. Of course, in the
case of Chess the game space is so huge that it is still unfeasible
today to calculate this theoretically optimal strategy, even with
the help of parallel processing in supercomputers. In contrast, Checkers
is a much smaller (3x3) and simpler game, making it possible to create
the complete game space in any typical desktop computer\footnote{In Checkers, the board size is 3x3 and each position can be either
empty or host the mark of of one of the two players, ``X'' or ``O''.
Hence, if the two players are treated as interchangeable (i.e., who
plays first) and no other symmetries are considered, the total number
of all possible distinct board setups is: $9\cdot8\cdot\ldots\cdot2\cdot1=9!=362,880$.
After applying the game rules and pruning the game tree for early
stops (with incomplete boards), the true number of game states is
about $\nicefrac{2}{3}$ of that set. Using simple tree-node representation
for each board setup, e.g. a 3-value 9-positions vector dictionary
($=3^{9}\simeq2^{14.265}\leq2^{15}<2^{16}=2$ bytes), such a program
would only require about 484 KB or less than 0.5 MB. This is roughly
the size of a small-sized photo taken by the camera of a low-end smart-phone
today, while in the '80s this was almost the total size of RAM in
a typical PC.} and calculate the exact optimal strategy - in fact, it is the same
strategy that every child soon learns by trial-and-error, playing
in a way that always leads to a win or a draw (never loose). 

In general, if the chosen strategy of one player is known to its opponent,
then an optimal counter-strategy is always available. Hence, in \emph{simultaneous}
games where the opposing moves are conducted at the same time, each
player would normally try not to employ a deterministic way of choosing
its strategy and conceal this choice until the very last moment. However,
the Minimax theorem provides a mathematically solid way of nullifying
any stochastic aspect in determining the opponent's choice and, in
essence, make its exact choice irrelevant: no matter what the opponent
does, the Minimax solution ensures the \emph{minimum losses} to each
player, given a specific game setup. In other words, it provides an
analytic way to determine the best \emph{defensive} strategy, instead
of a preference to offensive strategies. In some zero-sum games this
leads to one stable outcome or \emph{equilibrium}, where each player
would have no incentive not to choose its Minimax strategy; however,
if this choice leads to a negative handicap for this player if it
is known with complete certainty by the others, then this choice should
not be manifested as certain. In practice this means that the Minimax
solution would not be any single one of the player's \emph{pure} strategies
but rather a weighted combination of them in a \emph{mixed} strategy
scheme, where each weight corresponds to the probability of choosing
one of the available pure strategies via a random mechanism. This
notion of using mixtures of pure strategies for randomly choosing
between them leads to a false sense of security in single-turn games,
since the optimality of the expected outcome of the mixed strategy
scheme refers to the asymptotic (long-term) and not the ``spot''
(one-shot) payoff. Moreover, a game may involve an infinite number
of strategies for the players, in a discrete or continuous set; in
this case the game is labeled as \emph{continuous} or \emph{infinite},
while a \emph{finite} game is one with a limited number of (discrete)
strategies \cite{Dresher1961RAND,Thomas1984}.

When the game is inherently repetitive or \emph{iterative}, i.e.,
includes multiple turns and not just one, even the pure strategy suggested
by Minimax should not be chosen deterministically in every turn if
according to the game setup this information might provide a handicap
to the opponent. This is a topic of enthusiastic discussion about
the optimality of the Minimax solution and its inherent defensive
nature, as it is not clear in general when information about an opponent's
\emph{next} move is available and trustworthy enough to justify any
deviation from this Minimax strategy.

\parbox[c]{0.9\columnwidth}{%
\begin{shaded}%
\emph{Summary:}
\begin{itemize}
\item In \emph{zero-sum} games, one player's gains is another's losses (and
vice versa).
\item Information about the game structure and the opponents' moves may
be \emph{complete} or not, \emph{perfect} or not.
\item All board games are inherently zero-sum, of complete and perfect information.
\item The \emph{Minimax theorem} assures that all board games have at least
one theoretically optimal way to play them, although its exact calculation
may be unfeasible in practice for some games (e.g. Chess, Go).
\item The \emph{Minimax solution} of a game is the combination of players'
strategies that lead to an \emph{equilibrium} or \emph{saddle-point}.\end{itemize}
\end{shaded}%
}

\subsection{Nonzero-sum games and Nash equilibria}

Although the Minimax theorem provided a solid base for solving many
types of games, it is only applicable in practice for the zero-sum
type of games. In reality, it is common that in a conflict not all
players receive their opponents\textquoteright{} looses as their own
gain and vice versa. In other words, it is very common a specific
combination of decisions between the players to result in a certain
amount of \textquotedblleft loss\textquotedblright{} to one and a
corresponding \textquotedblleft gain\textquotedblright , not of equal
magnitude, to another. In this case, the game is called \emph{nonzero-sum}
and it requires a new set of rules for estimating optimal strategies
and solutions. As each player\textquoteright s gains and losses are
not directly related to the opponents\textquoteright , the optimal
solution is only based on the assertion that it should be the one
that ensures that the player has \textquotedblleft no regrets\textquotedblright{}
when choosing between possible decision options. This essentially
means that, since each player is now interested in his/her own gains
and losses, the optimal solution should only focus on maximizing each
player\textquoteright s own \emph{expectations} \cite{Owen1995,Montet2003,Dixit2008gt}.
The Minimax property can still be applied in principle when the single
most \textquotedblleft secure\textquotedblright{} option must be identified,
but now the solution of the game gains a new meaning.

During the early 1950\textquoteright s, John Nash has focused primarily
on the problem of finding a set of \emph{equilibrium points} in nonzero-sum
games, where the players eventually settle after a series of competitive
rounds of the game \cite{Nash1950,Nash1950phd}. The failure of the
Minimax approach to predict real-world outcomes in nonzero-sum games
comes from the fact that the players are assumed to act independently
and simultaneously, while in reality they usually are not. Experience
shows that \emph{possibly} better payoffs with what a player \emph{might}
choose, after observing the opponent's moves, is a very strong motivator
when choosing its actual strategy \cite{Mero1998}. In strict mathematical
terms, these equilibrium points would not be the same in essence with
the Minimax solutions, as they would come as a result of the players\textquoteright{}
competitive behavior over several ``turns'' of moves and not as
an algebraic solution of the mathematical formulation in a single-turn
game. 

In 1957 Nash has successfully proved that indeed such equilibrium
points exist in all nonzero-sum games, in a way that is analogous
to the Minimax theorem assertion. This new type of stable outcome
is referred to as \emph{Nash equilibrium} after his name and can be
considered a generalization of the corresponding Minimax equilibrium
in zero-sum games. In essence, they are the manifestation of the \emph{no
regrets} principle for all players, i.e., not regretting their final
choice after observing their opponents' behavior \cite{Stahl1999,Thomas1984}.
However, although the Nash theorem ensures that at least one such
Nash equilibrium exists in all nonzero-sum games, there is no clear
indication on how the game\textquoteright s solution can be analytically
calculated at this point. In other words, although a solution is known
to exist, there is no closed form for nonzero-sum games until today.
Seminal works by C. Daskalakis \& Ch. Papadimitriou in 2006-2007 and
on have proved that, while Nash equilibria exist, they may be unattainable
and/or practically impossible to calculate due to the inherent algorithmic
complexity of this problem, e.g. see: \cite{Daskalakis2006,Papadimitriou2011}.

It should be noted that players participating in a nonzero-sum game
may or may not have the same options available as alternative course
of action, or the same set of options may lead to different gains
or \emph{payoffs} between the players. When players are fully interchangeable
and their ordering in the game makes not difference to the game setup
and its solutions, the game is called \emph{symmetrical}. Otherwise,
if exchanging players\textquoteright{} position does not yield a proportional
exchange of their payoffs, then the game is called \emph{asymmetrical}.
Naturally, symmetrical games lead to Nash equilibrium points that
appear in pairs, as an exchange between players creates its symmetrical
counterpart.

\parbox[c]{0.9\columnwidth}{%
\begin{shaded}%
\emph{Summary:}
\begin{itemize}
\item In \emph{nonzero-sum} games, the payoffs of the players are separated
(although may be correlated).
\item If players are allowed to observe their opponents moves over several
iterations, then the ``no regrets'' principle is a strong incentive
to revise their own strategies, even though their payoffs are separated.
\item The \emph{Nash equilibrium theorem} ensures that, under these conditions,
there are indeed stable solutions in nonzero-sum games, similarly
to the Minimax theorem for zero-sum games.
\item However, calculating the optimal strategies and the game solution
for these Nash equilibria is a vastly more complex and generally unfeasible
task. \end{itemize}
\end{shaded}%
}

\section{Cooperation instead of competitiveness}

The seminal work of Nash and others in nonzero-sum games was a breakthrough
in understanding the outcome in real-world adversarial situations.
However, the Nash equilibrium points are not always the globally optimal
option for the players. In fact, the Nash equilibrium is optimal only
when players are strictly competitive, i.e., when there is no chance
for a mutually agreed solution that benefits them more. These strictly
competitive forms of games are called \emph{non-cooperative} games.
The alternative option, the one that allows communication and prior
arrangements between the players, is called a \emph{cooperative} game
and it is generally a much more complicated form of nonzero-sum gaming.
Naturally, there is no option of having cooperative zero-sum games,
since the game structure itself prohibits any other settlement between
the players other than the Minimax solution.

\subsection{The cooperative option}

The problem of cooperative or possibly cooperative gaming is the most
common form of conflict in real life situations. Since nonzero-sum
games have at least one equilibrium point when studied under the strictly
competitive form, Nash has extensively studied the cooperative option
as an extension to it. However, the possibility of finding and mutually
adopting a solution that is better for both players than the one suggested
by the Nash equilibrium, essentially involves a set of behavioral
rules regarding the players\textquoteright{} stance and \textquotedblleft mental\textquotedblright{}
state, rather than strict optimality procedures \cite{Mero1998}.
Nash named this process a \emph{bargain} between the players, trying
to mutually agree on one solution between multiple candidates within
a \emph{bargaining set }or\emph{ negotiation set}. In practice, each
player should enter a bargaining procedure if and only if there is
a chance that a cooperative solution exists and it provides at least
the same gain as the best strictly competitive solution, i.e., the
best Nash equilibrium. In this case, if such a solution is agreed
between the players, it is called \emph{bargaining solution} of the
game \cite{Montet2003,Owen1995}.

As mentioned earlier, each player acts upon the property of no regrets,
i.e., follow the decisions that maximize their own expectations. Nevertheless,
the game setup itself provides means of improving the final gain in
an agreed solution. In some cases, the bargaining process may involve
the option of \emph{threats}, that is a player may express the intention
to follow a strategy that is particularly costly for the opponent.
Of course, the opponent can do the same, focusing on a similar threat.
This procedure is still a cooperative bargaining process, with the
threshold of expectations raised for both players. The result of such
a process may be a mutually \emph{deterring} solution, which in this
case is called a \emph{threatening solution} or \emph{threat equilibrium}.
There is also evidence that, while cooperative strategies do exist,
in some cases ``cooperation'' may be the result of \emph{extortion}
between players with unbalanced power and choices \cite{WPress2012}.

In his work, Nash has formulated a general and fairly logical set
of six axioms, the \emph{Nash's bargaining axioms}, regarding the
behavior of \emph{rational} players, in order to establish a non-empty
bargaining set, i.e., to have at least one stable solution (equilibrium)
\cite{Montet2003,Owen1995,Nash1950}. In non-strict form, these axioms
can be summarized in the following propositions:
\begin{itemize}
\item Any of the cooperative options under consideration must be feasible
and yield at least the same payoff as the best strictly non-cooperative
option for all players, i.e., cooperation must be mutually beneficial. 
\item Strict (mathematical) constraints: Pareto optimality, independence
of irrelevant alternatives, invariance under linear transformations,
symmetry \cite{Thomas1984,Owen1995,Montet2003}.
\end{itemize}
The first proposition essentially defines the term ``rationality''
for a player: he/she always acts with the goal of maximizing own gains
and minimizing losses, regardless if this means strictly competitive
or possibly cooperative behavior. The second proposition names a set
of strict mathematical preconditions (not always satisfied in practice),
in order for such a bargaining set to exist. Having settled on these
axioms, Nash was able to prove the corresponding \emph{bargaining
theorem}: under these axioms, there exists such a \emph{bargaining
process}, it is unique and it leads to a bargaining solution, i.e.,
equilibrium. However, as in the general case of strictly competitive
games, Nash\textquoteright s bargaining theorem does not provide analytical
means of finding such solutions.

The notion of bargaining sets and threat equilibrium is often extended
in special forms of games that include iterative or recursive steps
in gaming, either in the form of multi-step analysis (\emph{meta-games})
or focusing on the transitional aspects of the game (\emph{differential
games}). Modern research is focused on methods that introduce probabilistic
models into games of multiple realizations and/or multiple stages
\cite{Owen1995}.

\parbox[c]{0.9\columnwidth}{%
\begin{shaded}%
\emph{Summary:}
\begin{itemize}
\item In nonzero-sum games, there may be non-competitive (cooperative) options
that are mutually beneficial to all players. 
\item Under some general rationality principles, \emph{Nash's bargaining
theorem} ensures that these cooperative outcomes may indeed become
the game solution, provided that strict competitiveness yields lower
gains for all.
\item The procedure of structuring the ``common ground'' of cooperation
between the players, normally conducted over several iterations, is
the bargaining process. \end{itemize}
\end{shaded}%
}

\subsection{Coalitions, stable sets, the Core}

Nash\textquoteright s work on the Nash equilibrium and bargaining
theorem provides the necessary means to study \emph{n}-person non-cooperative
and cooperative games under a unifying point of view. Specifically,
a nonzero-sum game can be realized as a strictly competitive or a
possibly cooperative form, according to the game\textquoteright s
rules and restrictions. Therefore, the cooperative option can be viewed
as a generalization to the strictly competitive mode of gaming.

When players are allowed to cooperate in order to agree on a mutually
beneficial solution of game, they essentially choose one strategy
over the others and bargain this option with all the others in order
to come to an agreement. For symmetrical games, this is like each
player chooses to join a group of other players with similar preference
over their initial choice. Each of these groups is called a \emph{coalition}
and it constitutes the basic module in this new type of gaming: the
members of each coalition act as cooperative players joined together
and at the same time each coalition competes over the others in order
to impose its own position and become the \emph{winning coalition}.
This setup is very common when modeling voting schemes, where the
group that captures the relative majority of the votes becomes the
winner.

Coalition Theory is closely related to the classical Game Theory,
especially the cooperating gaming \cite{Owen1995,Montet2003}. In
essence, each player still tries to maximize its own expectations,
not individually any more but instead as part of a greater opposing
term. Therefore, the individual gains and capabilities of each player
is now considered in close relation to the coalition this player belongs,
as well as how its individual decision to join or leave a coalition
affects this coalition\textquoteright s winning position. As in classic
nonzero-sum games, the notion of equilibrium points and solutions
is considered under the scope of domination or not in the game at
hand. Furthermore, the theoretical implications of having competing
coalitions of cooperative players is purely combinatorial in nature,
thus making its analysis very complex and cumbersome. There are also
special cases of collective decision schemes where a single player
is allowed to \emph{abstain} completely from the voting procedure,
or prohibit a contrary outcome of the group via a \emph{veto} option.

In order to study the properties of a single player participating
in a game of coalitions, it is necessary to analyze the wining conditions
of each coalition. Usually each player is assigned a fixed value of
\textquotedblleft importance\textquotedblright{} or \textquotedblleft weight\textquotedblright{}
when participating in this type of games and each coalition\textquoteright s
power is measured as a sum over the individual weights of all players
participating in this coalition. The coalition that ends up with the
highest cumulative value of power is the winning coalition. Therefore,
it is clear that, while each player\textquoteright s power is related
to its individual weight, this relation is \emph{not} directly mapped
on how the participation in any arbitrary coalition may affect this
coalition's winning or losing position. As this process stands true
for all possible coalitions that can be formed, this competitive type
of \textquotedblleft claiming\textquotedblright{} over the available
pool of players/voters by each coalition suggests that there are indeed
configurations that marginally favor the one or the other coalition,
i.e., a set of \textquotedblleft solutions\textquotedblright . 

The notion of solution in coalition games is somewhat different from
the one suggested for typical nonzero-sum games, as it identifies
minimal settings for coalitions that dominate all the others. In other
words, they do not identify points of maximal gain for a player or
even a coalition, but equilibrium ``points'' that determine which
of the forming coalitions is the winning one. This type of \textquotedblleft solutions\textquotedblright{}
in coalition games is defined in close relation to \emph{domination}
and \emph{stability} of such points and they are often referred to
as \emph{the Core}. Von Neumann and Morgenstern have defined a somewhat
more relaxed definition of such conditions and the corresponding solutions
are called \emph{stable sets} \cite{Owen1995,Montet2003}. It should
be noted that, in contrast to Nash\textquoteright s theorems and the
Minimax assertion of solutions, there is generally no guarantee that
solutions in the context of the Core and stable sets need to exist
in an arbitrary coalition game.

\parbox[c]{0.9\columnwidth}{%
\begin{shaded}%
\emph{Summary:}
\begin{itemize}
\item Players of similar preferences and mutual benefits may join in groups
or \emph{coalitions}; these coalitions may be competing with each
other, similarly to competitive games between single players. 
\item The study of games between coalitions is inherently more complex than
with single players, as in this case every player contributes to the
collective ``power'' and enjoys a share of the wins.
\item In general, coalitions are formed and structured under the scope of
\emph{voting ensembles}, where the voting weight of each individual
player contributes to the combined weight of the coalition.\end{itemize}
\end{shaded}%
}

\subsection{Indices of power in committees}

The notion of the Core and stable sets in coalition gaming is of vital
importance when trying to identify the winning conditions and the
relative power of each individual player in affecting the outcome
of the game. The observation that a player\textquoteright s weight
in a weighted system may not intuitively correspond to its voting
\textquotedblleft power\textquotedblright{} goes back at least to
Shapley and Shubik (1954). For example, a specific weight distribution
to the players may make them relatively equivalent in terms of voting
power, while only a slight variation of the weights may render some
of them completely irrelevant on determining the winning coalition
\cite{Taylor1993}. 

Shapley and Shubik (1954) and later Banzhaf and Coleman (1965, 1971)
suggested a set of well-defined equations for calculating the relative
power of each player, as well as each forming coalitions as a whole
\cite{Owen1995,Montet2003}. The \emph{Shapley-Shubik index of power}
is based on the calculation of the actual contribution of each player
entering a coalition, in terms of improving a coalition\textquoteright s
gain and winning position. Similarly, the \emph{Banzhaf-Coleman index
of power} calculates how an individual player\textquoteright s decision
to join or leave a coalition (``swing vote'') results in a winning
or loosing position for this coalition, accordingly. Both indexes
are basically means of translating each player\textquoteright s individual
importance or weight within the coalition game into a quantitative
measure of power in terms of determining the winner. While both indices
include combinatorial realizations, the Banzhaf index is usually easier
to calculate, as it is based on the sum of \textquotedblleft shifts\textquotedblright{}
on the winning condition a player can incur \cite{Berg1997}. Furthermore,
its importance in coalition games is made clearer when the Banzhaf
index is viewed as the direct result of calculating the derivatives
of a \emph{weighted majority game} (WMG).

Seminal work by L. S. Penrose \cite{Penrose1952}, as well as more
recent studies with computer simulations \cite{Chang2006}, have shown
that this discrepancy between voting weights and actual voting power
is clearly evident when there is large variance in the weighting profile
and/or when the voting group has less than 12-15 members. Even in
large voting pools, the task of designing optimal voting mechanisms
and protocols with regard to some \emph{collective efficiency} criterion
is one of the most challenging topics in Decision Theory. 

\parbox[c]{0.9\columnwidth}{%
\begin{shaded}%
\emph{Summary:}
\begin{itemize}
\item \emph{Weighted majority games} (WMG) are the typical theoretical structures
of the process of formulating the collective decision within a coalition.
\item In voting ensembles, each player's voting weight is \emph{not} directly
proportional to his/her true \emph{voting power} within the group,
i.e., the level of steering the collective decision towards its own
choices.\end{itemize}
\end{shaded}%
}

\subsection{Voting ensembles and majority winners}

In most cases, majority functions that are employed in practice very
simplistic when it comes to weighting distribution profile or they
imply a completely uniform weight distribution. However, a specific
weighting profile usually produces better results, provided that is
simple enough to be applied in practice and attain a consensus in
accepting it as \textquotedblleft fair\textquotedblright{} by the
voters. Taylor and Zwicker \cite{Taylor1993} have defined a voting
system as \emph{trade robust} if an arbitrary series of trades among
several winning coalitions can never simultaneously render them losing.
Furthermore, they proved that a voting system is trade robust if and
only if it is weighted. This means that, if appropriate weights are
applied, at least one winning coalition can benefit from this procedure. 

As an example, institutional policies usually apply a non-uniform
voting scheme when it comes to collective board decisions. This is
often referred to as the \textquotedblleft inner cabinet rule\textquotedblright .
In a hospital, senior staff members may attain increased voting power
or the chairman may hold the right of a tie-breaking vote. It has
been proven both in theory and in practice that such schemes are more
efficient than simple majority rules or any restricted versions of
them like trimmed means. Nitzan and Paroush \cite{Nitzan1982} have
studied the problem of optimal \emph{weighted majority rules} (WMR)
extensively and they have proved that they are indeed the optimal
decision rules for a group of decision makers in dichotomous choice
situations. This proof was later extended by Ben-Yashar and Paroush,
from dichotomous to polychotomous choice situations \cite{R.Ben-Yashar2001};
hence, the optimality of the WMR formulation has been proven theoretically
for any \emph{n}-label voting task. 

The weight optimization procedure has been applied experimentally
in trained or other types of combination rules, but analytical solutions
for the weights is not commonly used. However, Shapley and Grofman
\cite{Grofman1984} have established that an analytical solution for
the weighting profile exists and it is indeed related to the individual
player skill levels or \emph{competencies} \cite{Karotkin1998}. Specifically,
if decision independence is assumed for the participating players,
the optimal weights in a WMR scheme can be calculated as the log-odds
of their respective skill probabilities, i.e.:
\begin{equation}
w_{k}=\log\left(O_{k}\right)=\log\left(\frac{p_{k}}{1-p_{k}}\right)
\end{equation}
where $p_{k}$ is the competency of player \emph{k} and $w_{k}$ is
its corresponding voting weight. Interestingly enough, this is exactly
the solution found by analytical Bayesian-based approaches in the
context of decision fusion of independent experts in Machine Learning
\cite{Kunchev2004}. The optimality assertion regarding the WMR, together
with an analytical solution for the optimal weighting profile, provides
an extremely powerful tool for designing theoretically optimal collective
decision rules. Even when the independence assumption is only partially
satisfied in practice, studies have proved that WMR-based models employing
log-odds weighting profiles for combining pattern classifiers confirm
these theoretical results \cite{Georgiou2006,Georgiou2013}.

\parbox[c]{0.9\columnwidth}{%
\begin{shaded}%
\emph{Summary:}
\begin{itemize}
\item \emph{Weighted majority rules} (WMR) have been proven theoretically
as the optimal decision-making structures in weighted majority games.
\item The \emph{log-odds} model has been proven both as the theoretically
optimal way to weight the individual player's votes, provided that
they decide independently.
\item The optimality of the log-odds weighting method has also been proven
experimentally, even when the independence assumption is only partially
satisfied.\end{itemize}
\end{shaded}%
}

\subsection{Collective efficiency}

Condorcet (1785) \cite{Condorcet1989} was the first to address the
problem of how to design and evaluate an efficient voting system,
in terms of fairness among the people that participating in the voting
process, as well as the optimal outcome for the winner(s). This first
attempt to create a probabilistic model of a voting body is known
today as the \emph{Condorcet Jury Theorem} \cite{Young1988}. In essence,
this theorem says that if each of the voting individuals is somewhat
more likely than not to make the \textquotedblleft better\textquotedblright{}
choice from a set of alternative options; and if each individual makes
its own choice \emph{independently} from all the others, then the
probability that the group majority is \textquotedblleft correct\textquotedblright{}
is greater than the individual probabilities of the voters. Moreover,
this probability of correct choice by the group increases as the number
of independent voters increases. In practice, this means that if each
voter decides independently and performs marginally higher than 50\%,
then a group of such voters is \emph{guaranteed} to perform better
than each of the participating individuals. This assertion has been
used in Social sciences for decades as a proof that decentralized
decision making, like in a group of juries in a court, performs better
than centralized expertise, i.e., a sole judge. The Condorcet Jury
Theorem and its implications have been used as one guideline for estimating
the efficiency of any voting system and decision making in general
\cite{Young1988}. Under this context, the coalition games are studied
by applying quantitative measures on \emph{collective competence}
and \emph{optimal distribution of power} in the ensemble, e.g. tools
like the Banzhaf or Shapley indices of power. The degree of consistency
of such a voting scheme on establishing the pair-wise winner(s), as
the Condorcet Jury Theorem indicates, is often referred to as the
\emph{Condorcet criterion}.

Shapley-Shubik and Banzhaf-Coleman are only two of several formulations
for the indices of power in voting ensembles, each defining different
payoff distributions or realizations among the members of winning
coalitions. In general, these formulations are collectively referred
to as \emph{semivalue functions} or \emph{semivalues} and they are
considered more or less equivalent in principle, although may be different
in exact values. Almost all of them are based on combinatorial functions
(inclusion-exclusion operations in subsets) and, as a result, there
is no easy way to formulate proper inverse functions that can be calculated
in polynomial time. Therefore, the design of exact voting profiles
with weights based on semivalues, instead of competencies as described
above (log-odds), is generally impractical even for ensembles of small
sizes.

For further insight on weighted majority games, weighted majority
voting, collective decision efficiency and Condorcet efficiency, as
well as applications to Machine Learning for designing pattern classifiers,
see \cite{Georgiou2015,Georgiou2006,Georgiou2013}.

\parbox[c]{0.9\columnwidth}{%
\begin{shaded}%
\emph{Summary:}
\begin{itemize}
\item Under the assumption of independent voters and that each decides ``correctly''
marginally higher than 50\% of the time, then their collective decision
as a group is theoretically proven to be asymptotically better any
single member of the ensemble. 
\item Furthermore, as the size of the ensemble increases, its collective
competency is guaranteed to increase too.
\item In the other hand, the problem of formulating an analytical solution
for the optimal distribution of voting power within such a group,
i.e., the design of theoretically optimal \emph{voting mechanisms},
is still an open research topic.\end{itemize}
\end{shaded}%
}

\section{Game analysis \& solution concepts}

One of the most important factors in understanding and analyzing games
correctly is the way they are represented. Games can be represented
and analyzed in two generic formulations: (a) the \emph{analytical}
or \emph{normal} form, where each player is manifested as one dimension
and its available choices (strategies) as offsets on it, and (b) the
\emph{extensive} or \emph{tree-graph} form, where each player's ``move''
correspond to a node split in a tree representation. Each one of them
has its own advantages and disadvantages, but theoretically they are
equivalent.

\subsection{Games in analytical (matrix) form}

In Table \ref{tab:sample-game-analytical-matrix}, an example of a
zero-sum game in analytical form is presented. Player-1 is usually
referred to as the ``max'' player and Player-2 is referred to as
the ``min'' player, while rows and columns correspond to each player's
available strategies, respectively. Since this is a zero-sum game
and one player's gains is the other player's losses, the ``max''
player tries to maximize the game value (outcome) while the ``min''
player tries to minimize it. In the context of the Minimax theorem,
Player-2 chooses the \emph{maximum-of-minimums}, while Player-2 chooses
the \emph{minimum-of-maximums}. The \emph{x} and \emph{y} correspond
to the weight or probability of choosing the first strategy and, since
this is a 2x2 game, the other strategies are attributed with the complementary
probabilities, 1-\emph{x} and 1-\emph{y}. 

\begin{table}
\protect\caption{\label{tab:sample-game-analytical-matrix}Generic 2x2 zero-sum game
in analytical form.}

\centering{}%
\begin{tabular}{ccrccc}
\noalign{\vskip0.1cm}
\multirow{4}{0.1cm}{} & \multirow{2}{1.5cm}{\emph{Game example}} &  & \multicolumn{2}{c}{Player-2} & \multirow{4}{0.7cm}{}\tabularnewline
 &  &  & \emph{$y$} & \emph{$1-y$} & \tabularnewline
\cline{3-5} 
\noalign{\vskip0.1cm}
 & \multirow{2}{1.5cm}{Player-1} & \emph{$x$} & $a$ & $b$ & \tabularnewline
 &  & \emph{$1-x$} & $c$ & $d$ & \tabularnewline
\cline{3-5} 
\end{tabular}
\end{table}

The exact Minimax solution for \emph{x} and \emph{y} depends solely
on the values of the individual payoffs for each of the four outcomes.
Here, it is assumed that there is no \emph{domination} in strategies,
i.e., there is no row/column that is strictly ``better'' than another
row/column (column-wise/row-wise, respectively, all payoffs). For
example, Player-1 would have a dominating strategy in the first row
if and only if $a\geq c$ and $b\geq d$. Based on this generic setup,
this is a typical 2x2 system of linear equations and, if no domination
is present, its solution can be determined analytically as \cite{Stahl1999,Dresher1961RAND,Maschler2013gt}:
\begin{equation}
\left[x,1-x\right]=\left[\frac{d-c}{a-b-c+d},\frac{a-b}{a-b-c+d}\right]\label{eq:zerosum-maximin-solution}
\end{equation}
\begin{equation}
\left[y,1-y\right]=\left[\frac{d-b}{a-b-c+d},\frac{a-c}{a-b-c+d}\right]\label{eq:zerosum-minimax-solution}
\end{equation}
\begin{equation}
u=\frac{ad-bc}{a-b-c+d}\label{eq:zerosum-minimax-gamevalue}
\end{equation}

The Minimax solution $[x,y]$ determines the saddle-point, i.e., the
equilibrium that is reached when both opponents play optimally in
the Minimax sense, when the game has no pure (non-mixed) solution.
In this case, the expected payoff or \emph{value} of the game for
both players is calculated by $u$ (remember, this is a zero-sum game).
If the game has a pure solution, then it is determined as either 0
or 1 for each probability \emph{$x$} and \emph{$y$}. Table \ref{tab:sample-game-analytical-matrix-pure}
illustrates a zero-sum game and the corresponding pure Minimax solution,
by selecting the appropriate strategies for each player. In this case,
``max'' Player-1 chooses the the maximum \{1\} between the two minimum
values \{-3,1\} from its own two possible worst-case outcomes, while
``min'' Player-2 chooses the the minimum \{1\} between the two maximum
values \{4,1\} from its own two possible worst-case outcomes. Hence,
the pure solution {[}1,1{]} is the Minimax outcome.

\begin{table}
\protect\caption{\label{tab:sample-game-analytical-matrix-pure}Example 2x2 zero-sum
game in analytical form.}

\centering{}%
\begin{tabular}{ccrccc}
\noalign{\vskip0.1cm}
\multirow{4}{0.1cm}{} & \multirow{2}{1.5cm}{\emph{Game example}} &  & \multicolumn{2}{c}{Player-2} & \multirow{4}{0.7cm}{}\tabularnewline
 &  &  & (0) & (1) & \tabularnewline
\cline{3-5} 
\noalign{\vskip0.1cm}
 & \multirow{2}{1.5cm}{Player-1} & (0) & 0 & -3 & \tabularnewline
 &  & (1) & 4 & \textbf{1} & \tabularnewline
\cline{3-5} 
\end{tabular}
\end{table}

In nonzero-sum games, the analytical form is still a matrix, but now
the payoffs for each player are separate, as illustrated in Table
\ref{tab:sample-NZgame-analytical-matrix}. Here, since the payoffs
are separated, both players are treated as ``max'' and the Minimax
solution for each one is calculated by selecting the maximum-of-minimums
as described before for zero-sum games, focused solely on its own
payoffs from each value pair. 

\begin{table}
\protect\caption{\label{tab:sample-NZgame-analytical-matrix}Example of a 2x2 nonzero-sum
game in analytical form.}

\centering{}%
\begin{tabular}{ccrccc}
\noalign{\vskip0.1cm}
\multirow{4}{0.1cm}{} & \multirow{2}{1.5cm}{\emph{Game example}} &  & \multicolumn{2}{c}{Player-2} & \multirow{4}{0.7cm}{}\tabularnewline
 &  &  & \emph{$y$} & $1-y$ & \tabularnewline
\cline{3-5} 
\noalign{\vskip0.1cm}
 & \multirow{2}{1.5cm}{Player-1} & \emph{$x$} & ($a_{1}$,$a_{2}$) & ($b_{1}$,$b_{2}$) & \tabularnewline
 &  & $1-x$ & ($c_{1}$,$c_{2}$) & ($d_{1}$,$d_{2}$) & \tabularnewline
\cline{3-5} 
\end{tabular}
\end{table}

Although a (pure) Minimax solution can always be calculated for nonzero-sum
games, the exact Nash equilibrium solution is a non-trivial task that
cannot be solved analytically in the general case. However, pure Nash
equilibrium outcomes can be identified by locating any payoff pairs
$(z,w)$ such that \emph{$z$} is the maximum of its column and \emph{$w$}
is the maximum of its row. In other words, every row for Player-1
is scanned and every entry in it is compared to the values in the
same column, marking it if it is the maximum among them; the same
process is conducted for every column for Player-2, scanning each
value row-wise for its maximum; any payoff pair that has both values
marked as maximums is a Nash equilibrium in the game. Table \ref{tab:sample-NZgame-analytical-matrix-NEcalc}
illustrates such an example, where asterisk ({*}) marks the identified
max-values and the single Nash equilibrium for {[}\emph{A},\emph{B}{]}
at (2,4). Here, although the strategies are the same for both players,
their (separated) payoffs are not, hence the game is referred to as
\emph{asymmetric}. According to the \emph{oddness theorem} by Wilson
(1971), the Nash equilibria almost always appear in odd numbers \cite{Stahl1999,Owen1995},
at least for \emph{non-degenerate} games, where mixed strategies are
calculated upon \emph{k} linearly independent pure strategies.

\begin{table}
\protect\caption{\label{tab:sample-NZgame-analytical-matrix-NEcalc}Example of a 2x2
nonzero-sum game with one Nash equilibrium at {[}\emph{A},\emph{B}{]}:(2,4).}

\centering{}%
\begin{tabular}{ccrccc}
\noalign{\vskip0.1cm}
\multirow{4}{0.1cm}{} & \multirow{2}{1.5cm}{\emph{Game example}} &  & \multicolumn{2}{c}{Player-2} & \multirow{4}{0.7cm}{}\tabularnewline
 &  &  & \emph{A} & \emph{B} & \tabularnewline
\cline{3-5} 
\noalign{\vskip0.1cm}
 & \multirow{2}{1.5cm}{Player-1} & \emph{A} & (3,3) & (2{*},4{*}) & \tabularnewline
 &  & \emph{B} & (4{*},1) & (1,2{*}) & \tabularnewline
\cline{3-5} 
\end{tabular}
\end{table}

\parbox[c]{0.95\columnwidth}{%
\begin{shaded}%
\emph{Summary:}
\begin{itemize}
\item Game representation in \emph{analytical} form introduces a game matrix,
with row and column positions associated to the strategies available
to the players and contents associated to the corresponding payoffs.
\item Analytical-form representation introduces very convenient ways to
identify Minimax solutions and Nash equilibria in games.
\item However, they are appropriate mostly for 2-player simultaneous games,
since any other configuration cannot be fully illustrated. \end{itemize}
\end{shaded}%
}

\subsection{Games in extensive (tree-graph) form}

\begin{figure}
\begin{centering}
\includegraphics[width=6cm]{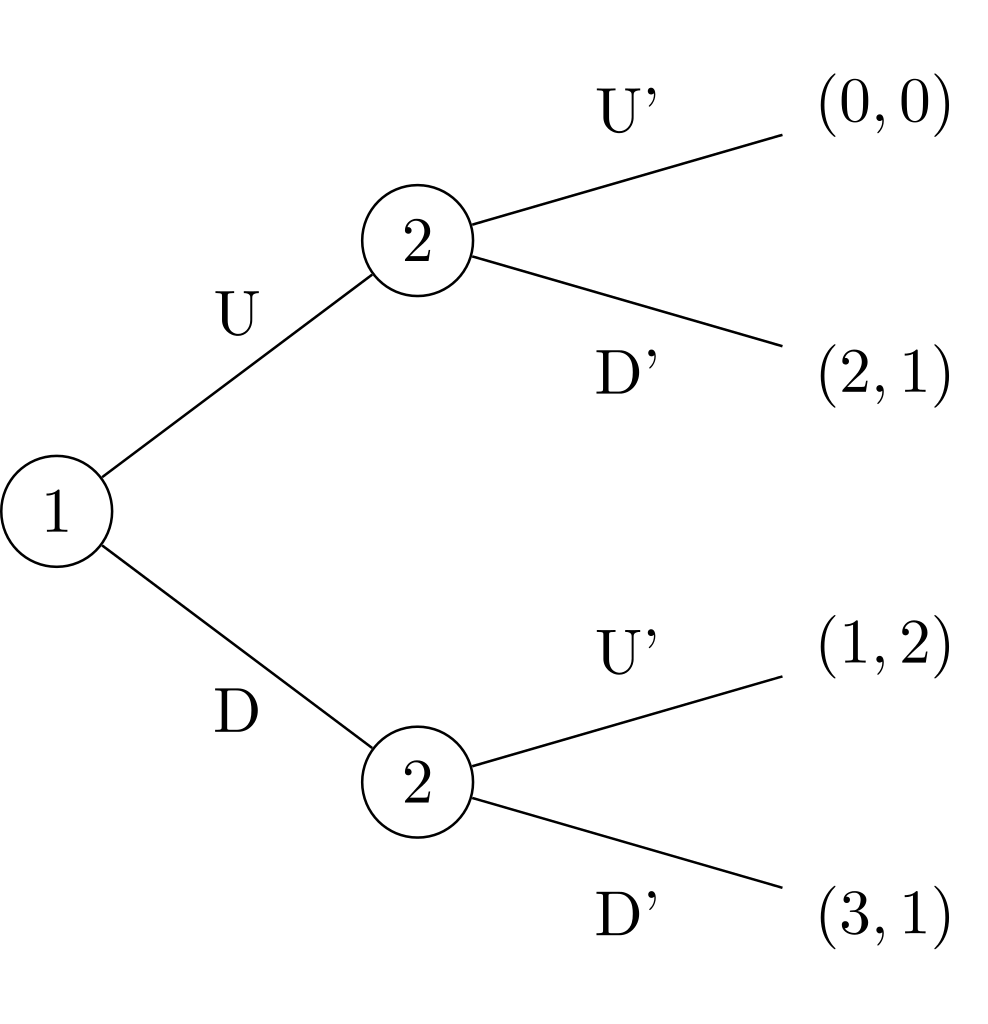}
\par\end{centering}

\protect\caption{\label{fig:sample-game-extensive-tree-perfectinfo}Example of a 2x2
nonzero-sum game of \emph{perfect} information.}
\end{figure}

\begin{figure}
\begin{centering}
\includegraphics[width=6cm]{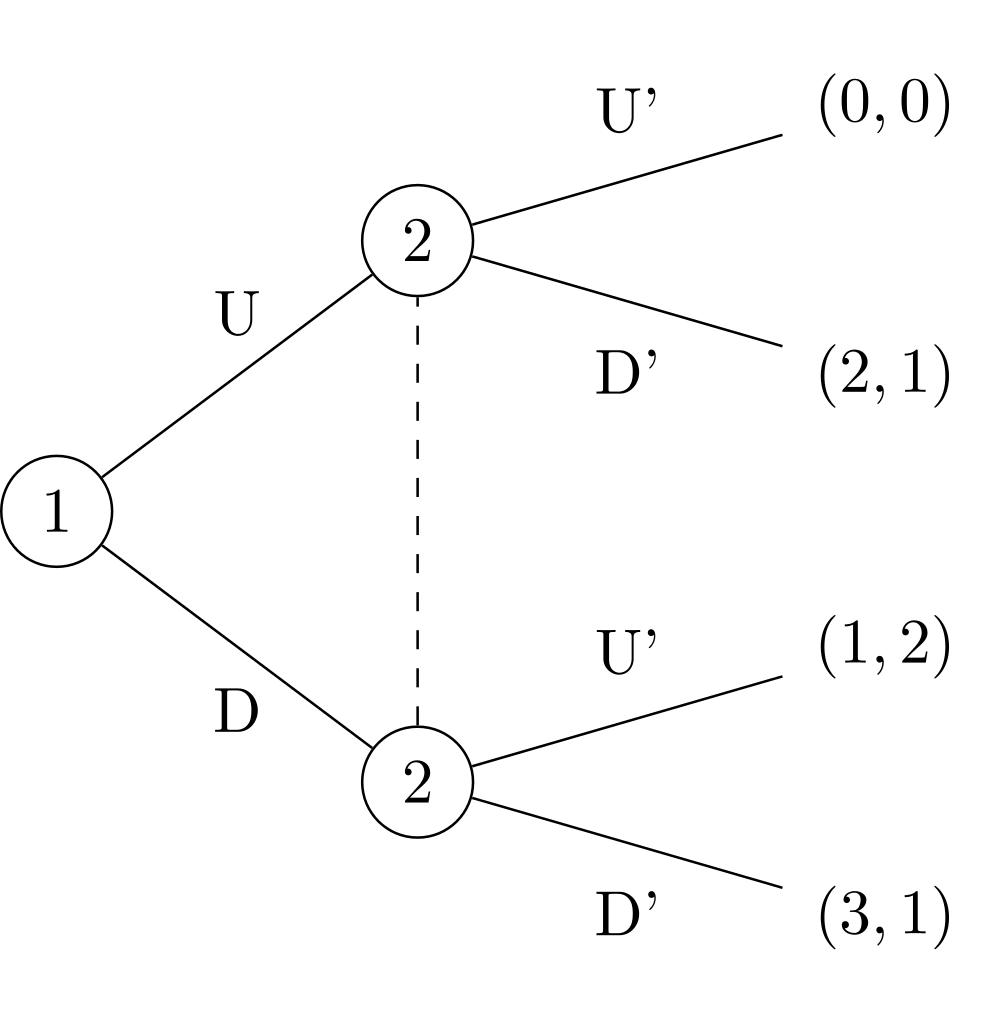}
\par\end{centering}

\protect\caption{\label{fig:sample-game-extensive-tree-uncertainty}Example of a 2x2
nonzero-sum game of \emph{imperfect} information.}
\end{figure}

In the extensive form the game is represented as a tree-graph, where
each node is a state labeled by a player's number and each (directed)
edge is a player's choice or ``move''. Strictly speaking, this is
a form of state-transition diagram that illustrates how the game evolves
as the players choose their strategies. Figure \ref{fig:sample-game-extensive-tree-perfectinfo}
shows such a 2x2 nonzero-sum game of perfect information, while Figure
\ref{fig:sample-game-extensive-tree-uncertainty} shows a similar
2x2 game of imperfect information \cite{Thomas1984,Montet2003,extformgame-wikiurl,Schalk2003,Tirole2003,Dresher1961RAND}.
Nodes with numbers indicate players, edges with letters indicate chosen
strategies (here, symmetric) and separated payoffs (in parentheses)
indicate the game outcome after one full round. The dashed line between
the two nodes for Player-2 indicate that its current true state is
not clearly defined due to imperfect information regarding Player-1's
move. In practice, these two states form an \emph{information set}
for Player-2, which has no additional information to differentiate
between them. This is also valid in the case of simultaneous moves,
where Player-2 cannot observe Player-1's move in advance of its own,
and vice versa. In extensive form, an information set is indicated
by a dotted line or by a loop, connecting all nodes in that set.

The extensive form of game is usually the preferred way to represent
the tree-graph of simple 2-player board games, where each node is
clearly a state and each edge is a player's move. Even in single-player
games, where a puzzle has to be solved through a series of moves (e.g.
Rubik's cube)\footnote{The combinatorial analysis of the classic 3x3x6 Rubik's cube should
take into account tile permutations that can only be reached by the
available shifts and turns of the slices of the device. Therefore,
a totally ``free'' permutation scheme would produce: $8!\cdot3^{8}\cdot12!\cdot2^{12}=519,024,039,293,878,272,000$
cube instances, while in practice the possible permutations are only:
$8!\cdot3^{7}\cdot\left(12!/2\right)\cdot2^{11}=43,252,003,274,489,856,000$
cube instances (about 12 times fewer) \cite{rubikscube-wikiurl}.}, the tree-graph is a very effective way to organize the game under
an algorithmic perspective, in order to program a ``solver'' in
a computer. In practice, the problem is structured as sequences of
states and transitions in a tree-graph manner and the ``game'' is
explored as it is evolving, move after move, expanding the tree-graph
from every terminal node. The tree-graph can be expanded either by
full a level (``breadth-first''), or from a branch all the way down
to non-expandable terminal nodes (``depth-first''), or some hybrid
scheme between these two alternatives. 

As described above, small games like Checkers can be structured and
expanded fully, with their tree-graph having only internal (already
expanded) and terminal nodes; however, in larger games like Chess
or Go this is practically unfeasible even with super-computers. In
such cases, the algorithm should assess the ``optimality'' of each
expandable terminal node with regard to relevance towards the predefined
goal (``win'' or ``solution''), sort all these nodes according
to their ranking and choose the ``best'' ones for expansion in the
next iteration. This way, the search is sub-optimal but totally feasible
with almost any memory constraints - this is exactly how most computer
players are programmed for playing board games or solving complex
puzzle games. In Artificial Intelligence, algorithms like \emph{A{*}}
and \emph{AB} solve this type of problems as a path-finding optimization
procedure towards a specified goal \cite{Russell2009,Nilsson1998}.

\begin{figure*}
\begin{centering}
\includegraphics[scale=0.55]{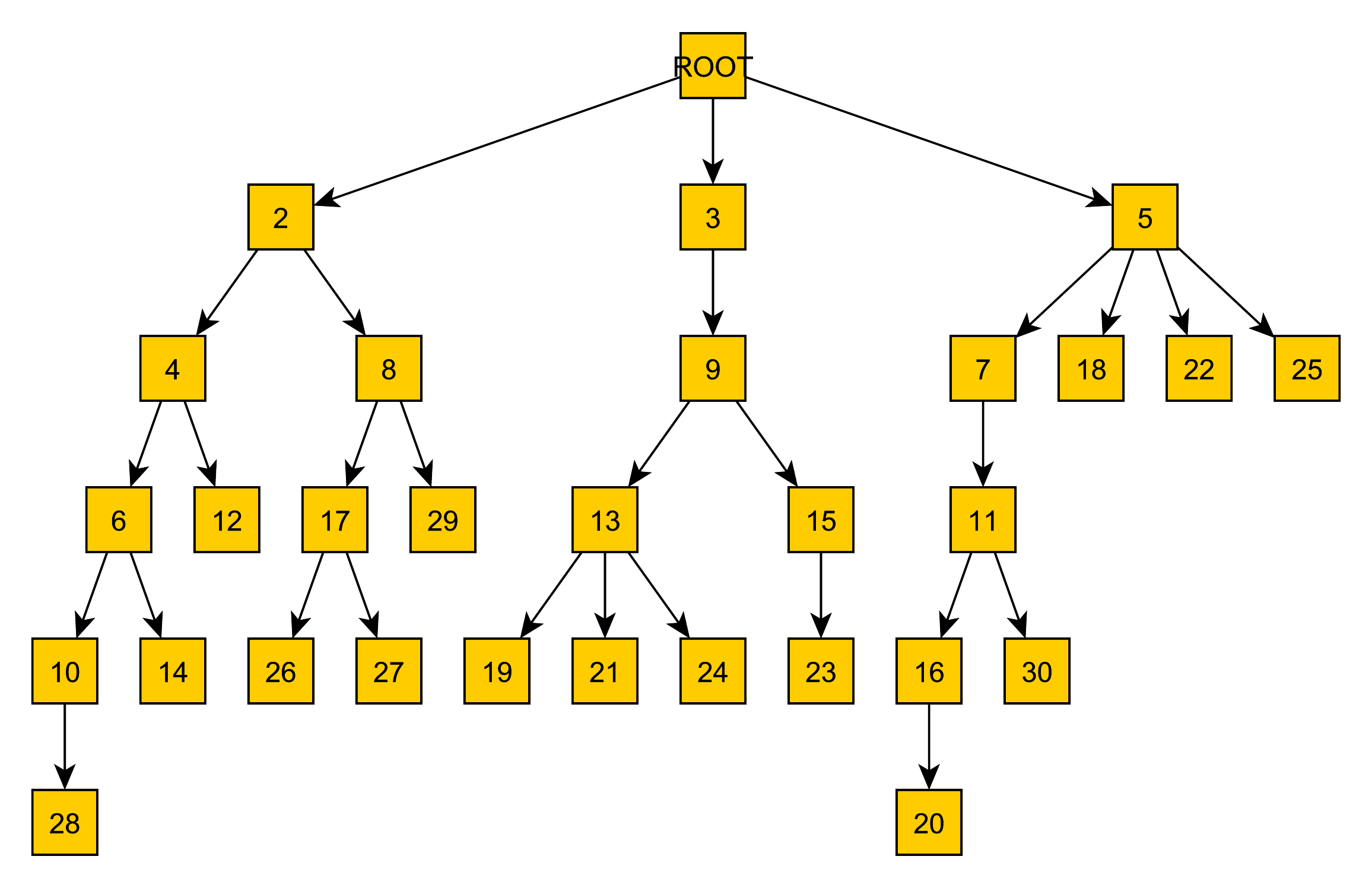}
\par\end{centering}

\protect\caption{\label{fig:sample-game-extensive-tree}Example of the way a path-finding
algorithm like A{*} would work in expanding the tree-graph of a single-player
``puzzle'' game like Rubik's cube.}
\end{figure*}

Figure \ref{fig:sample-game-extensive-tree} illustrates the way a
path-finding algorithm like A{*} would work in expanding a tree-graph
as described above. The ``root'' node is the starting state in a
puzzle game (single-player) and each node represents a new state after
a valid move. The numbers indicate the sequence in which the nodes
are expanded, according to some optimality-ranking function (not relevant
here). For example, node ``4'' in the 3rd level is expanded before
node ``5'' in the 2nd level, node ``21'' in the 5th level is expanded
before node ``22'' in the 3rd level, etc. Here, node ``30'' in
the 5th level is the last and most relevant terminal node (still expandable)
towards the goal, hence the optimal path from the ``root'' state
is currently the: ``5''$\rightarrow$``7''$\rightarrow$``11''$\rightarrow$``30''
and the next ``best'' single-step move is the one towards ``5''.
The tree-graph can be expanded in an arbitrary number of levels according
to the current memory constraints for the program, but the same path-finding
procedure has to be reset and re-applied after the realization of
each step when two or more players are involved, since every response
from the opponent effectively nullifies every other branch of the
tree-graph.

It should be mentioned that, although the extensive form of game representation
is often inefficient for large games like Chess, it can be used as
a tool in the proof of the existence of an optimal solution \cite{Ferguson2014gt,Thomas1984}.
Specifically, in every such game of complete and perfect information
(all board games), each player knows its exact position in the graph-tree
prior to choosing the next move. In other words, each player is not
only aware of the complete structure of the game but also knows all
the past moves of the game, including the ones of random choice. Hence,
since there is no uncertainty in the moves, each player can remove
the dominated strategies and subsequently identify the optimal choice,
which is always a pure strategy, i.e., the one that corresponds to
the saddle-point of the game. This proof actually ensures the existence
of a (pure) optimal strategy in every typical board game, no matter
how large or complex it is. Examples include Tic-Tac-Toe, Chess, Backgammon,
etc.

\parbox[c]{0.95\columnwidth}{%
\begin{shaded}%
\emph{Summary:}
\begin{itemize}
\item Game representation in \emph{extended} form introduces a tree-graph,
with nodes associated to individual players and (directed) edges associated
to selected strategies (``moves'').
\item Extended-form representation introduces very convenient ways to identify
chains of moves and solution paths.
\item However, the calculation of Minimax solutions and Nash equilibria
is not straight-forward. \end{itemize}
\end{shaded}%
}

\section{The four interesting cases}

In the real world, games may be either zero-sum or nonzero-sum by
nature. As described previously, the case of zero-sum games can be
considered simpler and much easier to solve analytically, since it
can be formulated as a typical algebraic set of linear equations that
define the Minimax solution, regardless if it contains pure or mixed
strategies \cite{Stahl1999,Dresher1961RAND}. However, nonzero-sum
games are inherently much more complex and require non-trivial solution
approaches, usually via some Linear Programming (constraint) optimization
procedure, e.g. see: \cite{Gu2008LP,Sierksma2001LP}. In fact, it
has been proven that the general task of finding the Nash equilibria
is algorithmically intractable\footnote{In their seminal works, Daskalakis, Goldberg and Papadimitriou have
shown that the task of finding a Nash equilibrium is PPAD-complete;
informally, PPAD is the class of all search problems which always
have a solution and whose proof is based on the parity argument for
directed graphs. Due to the proof of intractability, the existence
of Nash equilibrium in all nonzero-sum games somewhat loses its credibility
as a predictor of behavior.} \cite{Daskalakis2006,Daskalakis2009acm,Daskalakis2009siam,Papadimitriou2011}
- something that puts into a ``philosophical'' question the very
nature and practical usefulness of having proof of game solutions
(i.e., stable outcomes) that we may not be able to calculate.

Some cases of nonzero-sum games are particularly interesting, especially
when they involve symmetric configurations. The players can switch
places, the actual payoff values are usually of much less importance
than their relative ordering as a simple preference list, the Minimax
and Nash equilibria can be easily identified, yet these simple games
seem to capture the very essence of bargaining and strategic play
in a vast set of real-world conflict situations with no trivial outcomes.

\begin{table}
\protect\caption{\label{tab:NCG-general-matrix-form}The general analytical (matrix)
form of a 2x2 nonzero-sum symmetric game.}

\centering{}%
\begin{tabular}{ccrccc}
\noalign{\vskip0.1cm}
\multirow{4}{0.1cm}{} & \multirow{2}{1.5cm}{\emph{Game template}} &  & \multicolumn{2}{c}{Player-2} & \multirow{4}{0.7cm}{}\tabularnewline
 &  &  & \emph{C} & \emph{D} & \tabularnewline
\cline{3-5} 
\noalign{\vskip0.1cm}
 & \multirow{2}{1.5cm}{Player-1} & \emph{C} & (\emph{R,R}) & (\emph{S,T}) & \tabularnewline
 &  & \emph{D} & (\emph{T,S}) & (\emph{P,P}) & \tabularnewline
\cline{3-5} 
\end{tabular}
\end{table}

Table \ref{tab:NCG-general-matrix-form} shows a generic template
for such very simple symmetric nonzero-sum games, employing only two
strategies and four payoff values to completely define such games
in analytical (matrix) form. Here, the game is symmetric because the
players can switch roles without any effect in their corresponding
payoff pairs. Furthermore, they share two common strategies \emph{C}
and \emph{D}, named typically after the choices of ``cooperate''
or ``defect'', while constants \emph{P}, \emph{R}, \emph{S} and
\emph{T} are the real-valued payoffs in each case \cite{Casti1997}. 

In practice, a player's preference of strategies (and hence, the equilibria)
depends only on the relative ordering of the corresponding payoffs
and not their exact values, which become of real importance only when
the actual payoff value of the game solution is to be calculated for
each player. There is a finite number of rank combinations, i.e.,
permutations, of these four constants, which produce all the possible
unique game matrices of this type. Specifically, there are $4!=24$
different ways to order these four numbers, 12 of which can be discarded
as qualitatively equivalent to other game configurations. Out of the
12 remaining games, eight of them possess optimal pure strategies
for both players, therefore they can be considered trivial in terms
of calculating their solution. The four remaining configurations are
the most interesting ones, as they do not possess any optimal pure
strategy. These are the following:
\begin{itemize}
\item \emph{Leader}: $T>S>R>P$.
\item \emph{Battle of the Sexes}: $S>T>R>P$.
\item \emph{Chicken}: $T>R>S>P$.
\item \emph{Prisoner's Dilemma}: $T>R>P>S$.
\end{itemize}
These four qualitatively unique games seem to capture the essence
of most of the majority real-world conflict situations historically.
Although they have been studied extensively in the past, there are
still many open research topics regarding the feasibility, tractability
and stability of the theoretical solutions.

\subsection{Leader}

The \emph{Leader} or \emph{Coordination} game \cite{Montet2003,Owen1995,Thomas1984,Stahl1999,Casti1997,Tirole2003}
is named after the typical problem of two drivers attempting to enter
a stream of increased traffic from opposite sides of an intersection.
When the road is clear, each driver has to decide whether to move
in immediately or concede and wait for the other driver to move first.
If both drivers move in (i.e., choose \emph{D}), they risk crashing
onto each other, while if they both wait (i.e., choose \emph{C}),
they will waste time and possibly the opportunity to enter the traffic.
The former case is the worst, hence the payoff of (1,1), while the
later case is slightly more preferable with a payoff of (2,2). The
best outcome is for one driver to become the ``leader'' and move
first, while the other becomes the ``follower'' and move second.
There is still some difference in their absolute gains, but now the
deadlock is resolved in the best possible way, no matter who is actually
the leader and who is the follower.

Table \ref{tab:NCG-Leader-matrix-form} illustrates the analytical
form of this game setup, where numbers indicate relative preferences
rather than absolute gain values. There are two pure Nash equilibria,
(3,4) and (4,3), which correspond to the proper assignment of roles
to the players, explicitly or implicitly, such that coordination is
achieved. Since the game is symmetric the two players can switch roles,
with only marginal increase/decrease to their payoffs. In terms of
Minimax strategies, each player is free to choose the strategy that
guarantees the maximum-of-the-minimums without any concern about the
opponent's payoffs, since this is a nonzero-sum game. Hence, the Minimax
solution is {[}\emph{C},\emph{C}{]} at (2,2) marked in bold.

\begin{table}
\protect\caption{\label{tab:NCG-Leader-matrix-form}The typical setup of the \emph{Leader}
game with two players. Nash equilibria are marked with paired asterisks
and the Minimax solution with bold numbers.}

\centering{}%
\begin{tabular}{ccrccc}
\noalign{\vskip0.1cm}
\multirow{4}{0.1cm}{} & \multirow{2}{1.5cm}{\emph{Leader game}} &  & \multicolumn{2}{c}{Player-2} & \multirow{4}{0.7cm}{}\tabularnewline
 &  &  & \emph{C} & \emph{D} & \tabularnewline
\cline{3-5} 
\noalign{\vskip0.1cm}
 & \multirow{2}{1.5cm}{Player-1} & \emph{C} & (\textbf{2},\textbf{2}) & (3{*},4{*}) & \tabularnewline
 &  & \emph{D} & (4{*},3{*}) & (1,1) & \tabularnewline
\cline{3-5} 
\end{tabular}
\end{table}

In the real world, the assignment of roles as leader/follower is more
effective when applied explicitly, typically by some external mechanism
or a predefined set of rules. Street signs, traffic policemen and
highway code for driving properly are all such mechanisms for explicit
resolution of deadlocks via priority assignment in traffic.

\subsection{Battle of the Sexes}

In the \emph{Battle of the Sexes} game \cite{Montet2003,Owen1995,Thomas1984,Casti1997,Tirole2003},
a married couple has to decide between entertainment options for the
evening. The husband prefers one choice, while the wife prefers another.
The problem is that they would both prefer to concede to the same
choice together even if it is not their own, rather than follow their
own choices alone. For example, of he wants to watch a sports match
on TV and she wants to go out for dinner, they both prefer either
watching TV or going out for dinner as long as they are together.

Table \ref{tab:NCG-BattleSexes-matrix-form} illustrates the analytical
form of the game, where strategy \emph{C} is for conceding to the
other's preference and \emph{D} is for defecting to his/her own choice.
If they both concede the payoff (1,1) is the worst outcome, since
they both end up miserable and bored. If they both defect the payoff
(2,2) is marginally better for both, but they end up being alone.
The two other cases of someone following the other yields the best
payoffs for both, since the game is symmetric and they can switch
places. The outcomes (3,4) and (4,3) are actually the two Nash equilibria,
similarly to the \emph{Leader} game; however, the Minimax solution
(2,2) here corresponds to both players choosing \emph{D} (not \emph{C}
as in \emph{Leader}) as their best Minimax strategy.

\begin{table}
\protect\caption{\label{tab:NCG-BattleSexes-matrix-form}The typical setup of the \emph{Battle
of the Sexes} game with two players. Nash equilibria are marked with
paired asterisks and the Minimax solution with bold numbers.}

\centering{}%
\begin{tabular}{ccrccc}
\noalign{\vskip0.1cm}
\multirow{4}{0.1cm}{} & \multirow{2}{1.5cm}{\emph{Battle of the Sexes}} &  & \multicolumn{2}{c}{Player-2} & \multirow{4}{0.7cm}{}\tabularnewline
 &  &  & \emph{C} & \emph{D} & \tabularnewline
\cline{3-5} 
\noalign{\vskip0.1cm}
 & \multirow{2}{1.5cm}{Player-1} & \emph{C} & (1,1) & (3{*},4{*}) & \tabularnewline
 &  & \emph{D} & (4{*},3{*}) & (\textbf{2},\textbf{2}) & \tabularnewline
\cline{3-5} 
\end{tabular}
\end{table}

\subsection{Chicken}

One of the most well-known strategic games is \emph{Chicken} \cite{Ferguson2014gt,Maschler2013gt,Montet2003,Owen1995,Thomas1984,Casti1997},
dating back at least as far as the Homeric era. Two or more adversaries
engage in a very dangerous or even lethal confrontation, each having
a set of choices at his/her disposal and each of these choices producing
more or less damage to all players if their choice is the same. Typically,
this translates to the Hollywood's favorite version of two cars speeding
towards each other, the drivers can choose to turn and avoid collision
or keep the course and risk death if the other driver do not turn
either. The game seems simple enough, but there are several theoretical
implications that make it one of the most challenging situations,
appearing in many real-world conflicts throughout History.

Table \ref{tab:NCG-Chicken-matrix-form-1} illustrates the typical
\emph{Chicken} game setup with two players and two strategic choices.
Option \emph{C} corresponds to turning away (``swerve'') and losing
the game, while option \emph{D} corresponds to keeping the course
and risk death. The worst possible outcome is at (1,1) when players
persist in keeping course and eventually crashing against each other.
The mutually beneficial outcome or ``draw'' is at (3,3) when both
players decide to play safe and turn away; this is actually the Minimax
solution of the game, i.e., the most conservative and ``rational''
outcome if the game is a one-off round. On the other hand, there are
two Nash equilibria for the two outcomes when only one player turns
away and one persists.

\begin{table}
\protect\caption{\label{tab:NCG-Chicken-matrix-form-1}The typical setup of the \emph{Chicken}
game with two players. Nash equilibria are marked with paired asterisks
and the Minimax solution with bold numbers.}

\centering{}%
\begin{tabular}{ccrccc}
\noalign{\vskip0.1cm}
\multirow{4}{0.1cm}{} & \multirow{2}{1.5cm}{\emph{Chicken game}} &  & \multicolumn{2}{c}{Player-2} & \multirow{4}{0.7cm}{}\tabularnewline
 &  &  & \emph{C} & \emph{D} & \tabularnewline
\cline{3-5} 
\noalign{\vskip0.1cm}
 & \multirow{2}{1.5cm}{Player-1} & \emph{C} & (\textbf{3},\textbf{3}) & (2{*},4{*}) & \tabularnewline
 &  & \emph{D} & (4{*},2{*}) & (1,1) & \tabularnewline
\cline{3-5} 
\end{tabular}
\end{table}

One particularly interesting feature of the \emph{Chicken} game is
that it is impossible to avoid playing it with some insistent adversary,
since refusing to play is effectively equivalent to choosing \emph{C}
(swerve). Furthermore, the player who succeeds in making his/her commitment
to \emph{D} adequately convincing is always the one that can win at
the expense of the other player, assuming that the other player is
rational and would inevitably decide to avoid disaster. In other words,
the player that is somehow bounded to avoid losing \emph{at any cost}
and makes this commitment very clear to the opponent, is the one that
will always win against any rational player. 

This aspect of credible commitment is closely related to the notion
of \emph{reputation}, as well as the strange conclusion that in this
game the most effectively ``rational'' strategy is the manifestation
of ``irrational'' commitment to lethal risk. This becomes especially
relevant in cases where the game is played a number of times repeatedly
and previous behaviors directly affect the players' strategic choices
in the future: once the risky player starts winning he/she may maintain
or even improve this advantage, as confidence and prior ``risky''
behavior makes it more and more difficult for future opponents to
decide and deviate from their cautious Minimax choice of swerving.
The \emph{Chicken} game is perhaps the most descriptive and simple
case where players' previous behavior (i.e., \emph{reputation}) is
of such importance for predicting the actual outcome.

\subsection{Prisoner's Dilemma}

This forth basic type of non-trivial, nonzero-sum game is by far the
most interesting one. The \emph{Prisoner's Dilemma} game \cite{Ferguson2014gt,Maschler2013gt,Montet2003,Owen1995,Thomas1984,Stahl1999,Casti1997,Tirole2003}
typically involves two prisoners who are accused of a crime. Each
of them has the option of remaining silent and withholding any information
or confessing to the police and accusing the other by disclosing details
about the crime. The first choice \emph{C} is effectively the cooperative
option, while the second choice \emph{D} corresponds to purely competitive
behavior in order to reduce he/her own damages. 

Table \ref{tab:NCG-PrisonerD-matrix-form} illustrates the typical
\emph{Prisoner's Dilemma} game setup with two players and two strategic
choices. The payoffs here correspond simply to preferences and not
real gain/cost values, but the essence and the strategic properties
of the game remain intact. In practice, what the game matrix says
is that if the two prisoner's remain silent, i.e., mutually cooperate,
they will not be freed but they will share an equal, relatively mild
conviction. If they both talk and accuse each other, i.e., mutually
defect, they will share and equal but more severe conviction. If only
one of them talks to the police and the other remains silent, the
one that talked is freed and the other serves a full-time conviction
for both. It is of course imperative that the two prisoners are immediately
separated upon capture and no communication between them is allowed;
this does not nullifies any prior arrangements they may have, but
isolation after being captured means that neither of them can confirm
they loyalty of the other. This is one of the main reasons why police
always isolates suspects prior and during any similar investigation.

\begin{table}
\protect\caption{\label{tab:NCG-PrisonerD-matrix-form}The typical setup of the \emph{Prisoner's
Dilemma} game with two players. Nash equilibria are marked with paired
asterisks and the Minimax solution with bold numbers.}

\centering{}%
\begin{tabular}{ccrccc}
\noalign{\vskip0.1cm}
\multirow{4}{0.1cm}{} & \multirow{2}{1.5cm}{\emph{Prisoner's Dilemma}} &  & \multicolumn{2}{c}{Player-2} & \multirow{4}{0.7cm}{}\tabularnewline
 &  &  & \emph{C} & \emph{D} & \tabularnewline
\cline{3-5} 
\noalign{\vskip0.1cm}
 & \multirow{2}{1.5cm}{Player-1} & \emph{C} & (3,3) & (1,4{*}) & \tabularnewline
 &  & \emph{D} & (4{*},1) & (\textbf{2}{*},\textbf{2}{*}) & \tabularnewline
\cline{3-5} 
\end{tabular}
\end{table}

The real beauty and singularity of the \emph{Prisoner's Dilemma} is
that it implies a paradox. A quick analysis of the payoffs in Table
\ref{tab:NCG-PrisonerD-matrix-form} yields two extremes at (1,4)
and (4,1), corresponding to the two interchangeable cases one player
cooperating (\emph{C}) and one not (\emph{D}), but in contrast to
the three previous games these are \emph{not} Nash equilibria. There
is only one Nash equilibrium at (2,2), which is in fact the Minimax
solution too. This means that under the solution concepts of both
Minimax strategy and Nash equilibrium, theory suggests that the two
prisoner's will probably choose to betray one another, despite any
previous arrangements. It is clearly evident that the outcome (3,3)
is \emph{mutually beneficial} and at the same time unattainable due
to lack of communication. However, in therms of strict personal gain,
defecting (\emph{D}) is the dominant strategy for both and neither
of them has any incentive to deviate from it. In other words, it appears
that defecting is always the optimal choice \emph{regardless} of what
the other prisoner does - but if both adopt the same rationale, they
will end up at (2,2) which is clearly worse than the (3,3) that they
could have gotten if they had chosen mutual cooperation.

The essence of the paradox of \emph{Prisoner's Dilemma} lies in the
inherent conflict between \emph{individual }and\emph{ collective rationality}.
While individual rationality is well-understood, collective rationality
deals with the scope of optimizing the \emph{mutual} gain of the players.
This is not a default behavior in strictly competitive situations,
as in zero-sum games, or nonzero-sum games that do not imply cooperation.
However, nonzero-sum games permit the idea of \emph{mutually} optimal
gains as a combination of \emph{simultaneously} optimal separate payoffs.
Under this broader scope, even (4,1) and (1,4) are worse than (3,3)
since they yield a sum of 5 in gain value rather than 6, respectively.

It should also be noted that the single Nash equilibrium in \emph{Prisoner's
Dilemma} is stable, while the corresponding pairs of Nash equilibria
in the three previous games are inherently unstable, since the players
are not in agreement as to which of the two equilibria is preferable.
Furthermore, in the three previous games the worst possible outcome
comes when both players choose their non-Minimax strategy; in \emph{Prisoner's
Dilemma} this is not so. In fact, \emph{Prisoner's Dilemma} has produced
lengthy academic debates and hundreds of studies in a wide range of
disciplines, from Game Theory and Mathematics to Sociology and Evolutionary
Biology. The paradox of this game (as described above) has been illustrated
as a notorious example where theory often fails to predict the true
``gaming'' outcomes in the real world: cooperation can emerge spontaneously,
even though theory says it should not \cite{Axelrod1984,Axelrod1988,Mero1998,Casti1997}.

\parbox[c]{0.95\columnwidth}{%
\begin{shaded}%
\emph{Summary:}
\begin{itemize}
\item There are four basic nonzero-sum game types of particular interest
namely: \emph{Leader} (or \emph{Coordination}), \emph{Battle of the
Sexes}, \emph{Chicken} and \emph{Prisoner's Dilemma}. 
\item Three of these games (except \emph{Prisoner's Dilemma}) have two ``mirrored''
pure Nash equilibria and players receive the worst possible payoff
when they choose to deviate from their optimal Minimax strategy.
\item \emph{Prisoner's Dilemma} is a very unique type of game, since neither
Minimax solution or Nash equilibrium (single one in this case) point
to the best \emph{mutually} beneficial outcome; this is informally
labeled as the \emph{paradox} of this game.\end{itemize}
\end{shaded}%
}

\section{Signals, mechanisms \& rationality}

Game formulation and representation in analytical or extensive form
are imperative for proper analysis and identification of equilibria.
However, they fail to capture many elements of \emph{gaming} as a
multi-aspect process, especially in relation to \emph{strategic moves};
these are actions performed by the players at different places and
times, even before the realization of the current game, with the goal
of enhancing strategic advantages and increasing the effectiveness
of chosen strategies. Sometimes the ``moves'' are no more than message
exchanges between the players, explicit or implicit, or simply tracking
the history of previous choices in iterated games. Formulating these
factors into a proper mathematical model can be very difficult, but
nevertheless they are matters of great importance in real-world conflict
situations.

\subsection{Signals, carriers \& bluffs}

The exchange of messages between the players is a very useful option
when a player is trying to model or even predict the behavior of its
opponent(s). A message or \emph{signal} from one player to another
may be voluntary or involuntary, direct or indirect, explicit or implicit
\cite{Thomas1984,Stahl1999}. In any case, it carries some sort of
strategic information, which is always valuable to the other player
if it can be asserted as credible with a high degree of confidence.
On the other hand, if this credibility can be manipulated and falsely
asserted as such, the source player may gain some strategic advantage
by means of deceiving its opponent.

Strategic \emph{signaling} is the process of information exchange
between two or more players in a game, using any means or intermediate
third-parties as carriers. If the source player does this deliberately,
the purpose is to project some strategic preference or \emph{stance}
(``posturing'') in the game without making any actual ``move'',
in order to intimidate or coordinate with the opponent(s). This is
particularly useful in situations where mutually beneficial equilibria
are achievable but lack of preference ranking can lead to disastrous
lack of coordination. The \emph{Leader} and \emph{Battle of the Sexes}
games are such examples (see Tables \ref{tab:NCG-Leader-matrix-form}
and \ref{tab:NCG-BattleSexes-matrix-form}). On the other hand, if
the source player signals its opponent unintentionally, this strategic
information could be a ``leak'' of such importance that may determine
the actual outcome of the game.

\emph{Explicit} signaling means that the source player sends out a
clear message with undeniable association and content. An explicit
signal may be voluntary or involuntary; in the later case, the message
is simply a ``leak'' with very clear origin and content. \emph{Implicit}
signaling happens when the origin or (most commonly) the content of
the message is somehow inconclusive or ``plausibly deniable'' as
to the intentions of the source player. A signal exchange may occur
directly between the players or via a third-party that performs the
role of a carrier. A number of combinations of these attributes are
possible in practice, employing direct/indirect messaging, voluntary/involuntary
information exchange, with explicit/implicit messages. For example,
a third-party carrier may share an implicit signal or ``leaked''
(involuntary) information about a player's stance with another player,
participating in the game only as a mediator, coordinator or ``referee'',
rather than an actively involved player.

A very special type of signaling is when the message exchange involves
false information, i.e., a \emph{bluff}. This kind of signals is a
very common practice in games of imperfect and/or incomplete information,
where the players do not have a complete view of the game structure
itself and/or the opponents' choices, respectively. In this case,
\emph{false signaling} or \emph{bluffing} is usually a strategic option
by itself, exploiting this uncertainty regarding the true status of
the game to enhance advantages or mitigate disadvantages. A very common
example of such games is Poker, where a player with weaker deck of
cards can project a \emph{false stance} to its opponents, in order
to avoid defeat or even secure a victory against players with better
decks of cards \cite{Thomas1984,Stahl1999}. Bluffing can be realized
directly between players or indirectly via a third-party carrier.
In the later case, especially when the signaling is implicit and assumed
involuntary, the credibility of the assertion is strongly associated
with the credibility of the carrier itself. In other words, even if
the source player could not project a successful bluff on its own,
a credible third-party carrier might be the necessary intermediate
to achieve such a move. The role of third-party mediators in signaling
is a special topic in the study of strategic moves and how they affect
the final outcome in games.

\parbox[c]{0.95\columnwidth}{%
\begin{shaded}%
\emph{Summary:}
\begin{itemize}
\item A \emph{signal} between players is a voluntary or involuntary, direct
or indirect, explicit or implicit exchange of a message; it is usually
a declaration of \emph{stance} (``posture'') in the game, i.e.,
intent to include or exclude a strategy from a set of open options.
\item \emph{Strategic moves}, e.g. signaling, project some strategic preference
without making any actual ``move'', in order to intimidate or coordinate
with the other player(s).
\item A \emph{bluff} is a projection of false information, i.e., exploiting
the incomplete/imperfect information structure of a game to gain some
strategic advantage that could not be achievable if the game was of
complete/perfect information.\end{itemize}
\end{shaded}%
}

\subsection{Credibility, reputation, promises \& threats}

The effectiveness of projecting a strategic stance via signaling,
regardless if it is true or bluff, depends heavily on the \emph{credibility}
of that signal, as well as the credibility of the player itself \cite{Thomas1984,Stahl1999}.
When it comes to a single signal or stance, the credibility is closely
linked to the level of compatibility of that signal or stance with
the rationality of the player. Although rationality per se may be
only an assumption with regard to one's opponent, in general terms
it is fairly easy to examine the matrix or the tree-graph representation
of a game and establish whether a declared stance is beneficial or
not to the associated player. In other words, if that player is assumed
to behave rationally, Minimax strategies and Nash equilibria can be
used to filter out choices that are clearly excluded, at least with
a high probability. 

The set of previous stances and/or moves, as well as their associated
credibility values, can be used as the history or \emph{reputation}
of that player, which is in fact the a priori probability for any
future stance and/or move of being consistent with its previous behavior
\cite{Mero1998}. Since games of complete and perfect information,
e.g. Chess, are not compatible with false signaling and bluffs, the
true theoretical aspect of credibility and reputation is relevant
only in games of incomplete and/or imperfect information. Hence, Poker
players are indeed characterized as being cautious or risk-takers
according to their reputation on using bluffs in lower or higher frequency,
respectively.

A player with a specific reputation can signal a specific stance to
the others, projecting either a \emph{promise} or a \emph{threat}.
A promise is a signal that usually declares the intent to cooperate,
i.e., choose the less aggressive approach. This is particularly useful
when the players need to coordinate in order to avoid much worse outcomes,
as in the games \emph{Leader} and \emph{Battle of the Sexes} (see
Tables \ref{tab:NCG-Leader-matrix-form} and \ref{tab:NCG-BattleSexes-matrix-form}).
On the other hand, a threat is a signal that usually declares the
intent to compete, i.e., choose the more aggressive approach. This
is still useful as the means to enforce some kind of coordination,
now in the form of extortion rather than willful cooperation. The
\emph{Chicken} game is such any example (see Table \ref{tab:NCG-Chicken-matrix-form-1}),
where one player must force the other to swerve, in order to naturally
end up in one of the two Nash equilibria and avoid the worst outcome
of crash.

As it was mentioned earlier, \emph{Prisoner's Dilemma} is a very special
type of game, since neither Minimax solution or Nash equilibrium points
to the mutually beneficial option of cooperation; however, if signaling
between the prisoners is possible, i.e., if they are allowed to communicate
with each other, cooperation becomes much more plausible: all they
have to do is to promise each other to remain silent and threat to
accuse the other as a retaliation if they see the other doing such
thing. One of the most interesting topics in modern Game Theory is
the study and analytical formulation of the conditions, the constraints
and the exact processes of the \emph{evolution of cooperation} in
games like \emph{Prisoner's Dilemma}, where typical theory fails to
predict optimal strategies, although such strategies seem to exist,
usually in accordance to some \emph{Tit-for-Tat} variation \cite{Axelrod1984,Axelrod1988,Mero1998,Casti1997}.

In any case, whether it is a promise or a threat, the signal or stance
is labeled as credible or not. Hence, a \emph{credible promise} is
one that comes from a player with a reputation of being consistently
reliable in fulfilling that promise, i.e., actually choosing less
aggressive strategies when signaling intent to cooperate. Similarly,
a \emph{credible threat} is one that comes from a player with a reputation
of being consistently reliable in fulfilling that threat, i.e., actually
choosing more aggressive strategies when signaling intent to compete
\cite{Montet2003,Owen1995}.

\parbox[c]{0.95\columnwidth}{%
\begin{shaded}%
\emph{Summary:}
\begin{itemize}
\item \emph{Promise} is a signal that usually declares the intent to cooperate,
i.e., choose the less aggressive approach; it is useful when players
need to coordinate in order to avoid much worse outcomes.
\item \emph{Threat} is a signal that usually declares the intent to compete,
i.e., choose the more aggressive approach; it is useful a player wants
to enforce some kind of coordination, in the form of extortion.
\item \emph{Credibility} is closely linked to the level of compatibility
of a signal or stance with the rationality of the player; in practice,
it is a measure (probability) of whether the player will fulfill a
promise or a threat, if necessary.
\item \emph{Reputation} of a player is the a priori probability for any
future stance and/or move of being consistent with its previous behavior.
\item \emph{Credible promises} and \emph{credible threats} are associated
with the reputation and credibility of each player, as well as the
actual payoffs in the corresponding game matrix.\end{itemize}
\end{shaded}%
}

\subsection{Utility, incentives \& ``rational irrationality''}

As it was mentioned earlier, if that player is assumed to behave rationally,
i.e., trying to minimize losses and maximize gains in terms of actual
payoffs in each outcome, the credibility of a promise or a threat
can be easily established with a high probability. Nevertheless, the
fact that this is just a probability and not a perfect forecast comes
from the fact that, in turn, the level of rationality of that player
cannot be evaluated perfectly and in exact terms. 

Rationality and \emph{incentives} of a player emerge naturally from
the exact formulation of its own \emph{utility function}, which is
nothing more than a generalization of the loss/gain function that
is described by the matrix or the tree-graph of the game \cite{Owen1995,Tirole2003,Montet2003}.
If the formulation of the game's payoff matrix is perfect, then it
is clear when a strategy is optimal for a player and when it is not.
However, the truth is that these payoff values may not reflect the
exact \emph{utility}, i.e., overall loss/gain value for that player,
usually due to some ``hidden'' outcomes or side-effects. For example,
a game matrix may describe the payoffs for each outcome and each player
correctly, but with the assumption that these players are rational
in the same way: winning over their opponent; this may not be true,
e.g. when one player cares more about securing that their opponent
does not win, rather than securing their own win. In other words,
when the players' rationality is not symmetrically the same, then
they do not share the same utility function and the true payoffs in
the game matrix may actually be quite different. 

A very classic example of such games, assumed to be symmetric when
they are actually asymmetric by nature, is the \emph{Hostage Situation},
described in analytical form by Table \ref{tab:NCG-Hostage-matrix-form}.
If the two opponents are treated as similarly rational, i.e., symmetric
in terms of incentives and behavior, then the game is not much different
than the classic \emph{Chicken}, where one must convince the other
to swerve first, in order to avoid the crash. This translates to either
the authorities give in to the assaulter's demands or the assaulter
eventually surrenders to the authorities, both outcomes assumed to
be equally rational, correspondingly, to each player. However, if
for some reason the assaulter is more determined than initially presumed,
preferring to fight to the death rather than surrendering and ending
up in jail, then the game is inherently asymmetric and the payoff
matrix is quite different, as illustrated in Table \ref{tab:NCG-Hostage-matrix-form}.
What the matrix shows is that now Player-1, i.e., the assaulter, has
a dominant strategy of always choosing the most aggressive stance,
no matter what the authorities choose to do. There is no pure Minimax
solution here, since there is no pure saddle-point (see payoffs ``3''
and ``2'' in bold); however, there is now a single Nash equilibrium
at (4,2), i.e., aggressive assaulter and passive authorities - this
is in fact the standard approach internationally in all hostage situations:
the authorities start with trying to establish a communication link
and negotiate with the assaulter, rather than choosing a rescue operation
by direct action that could put the hostages in danger.

\begin{table}
\protect\caption{\label{tab:NCG-Hostage-matrix-form}The typical setup of the \emph{Hostage
Situation} game with two players. Player-1 is the assaulter and Player-2
is the rescuer-protector.}

\centering{}%
\begin{tabular}{ccrccc}
\noalign{\vskip0.1cm}
\multirow{4}{0.1cm}{} & \multirow{2}{1.5cm}{\emph{Hostage Situation}} &  & \multicolumn{2}{c}{Player-2} & \multirow{4}{0.7cm}{}\tabularnewline
 &  &  & \emph{C} & \emph{D} & \tabularnewline
\cline{3-5} 
\noalign{\vskip0.1cm}
 & \multirow{2}{1.5cm}{Player-1} & \emph{C} & (2,3) & (1,4{*}) & \tabularnewline
 &  & \emph{D} & (4{*},\textbf{2}{*}) & (\textbf{3}{*},1) & \tabularnewline
\cline{3-5} 
\end{tabular}
\end{table}

As it is evident from the \emph{Hostage Situation} game of Table \ref{tab:NCG-Hostage-matrix-form},
the authorities are normally guided to a more passive and cooperative
approach of negotiating rather than using force, because the incentive
is to protect the hostages at all costs. This effectively translates
to employing a utility function that includes a high priority on the
hostages' lives, higher than the immediate capture or incapacitation
of the assaulter. Hence, the rationality of Player-2 dictates a more
passive, cooperative stance. This changes drastically if, during this
evolution, the lives of hostages are put in severe danger, e.g. when
the assaulter poses a very credible threat or actually harms a hostage
(assuming there are more). In this case, the authorities should change
stance and employ the more aggressive option, because this is now
the optimal response.

Table \ref{tab:NCG-Kamikaze-matrix-form-1} illustrates the \emph{Kamikaze}
game, which is actually a slightly modified \emph{Hostage Situation}
game in terms of payoff matrix. The game is still asymmetric and the
only variation is the swapping of payoff values \{2\} and \{1\} for
Player-2 (marked in italics), which illustrates the new fact that
at this point it is more harmful for the hostages to remain idle rather
than using direct force to rescue them, even if this too poses some
danger to them - again, this is exactly the standard approach internationally
in all hostage situations: the authorities follow strict rules-of-engagement
which state that, once it is established that the lives of hostages
is in clear and severe danger, direct action is to be employed immediately.
The same setup emerges when the \emph{Kamikaze} game is studied according
to its name: when one player (assaulter) is more concerned about damaging
the opponent (defender) rather than protecting itself, then there
is indeed a dominant strategy of always choosing the most aggressive
stance, no matter what the defender chooses to do. Likewise, the defender
is now forced to choose between its two worst outcomes and naturally
chooses the less damaging one, i.e., direct counter-action rather
than swerve. Here, the passive stance is established as more damaging
than all-out-conflict, exactly as in \emph{Hostage Situation} with
a very aggressive assaulter. In terms of game analysis, now there
is indeed a pure Minimax solution at (3,2), which is also the single
Nash equilibrium of the game. This explains why there is practically
no other rational (strategically optimal) way to defend against a
murderous hostage-taker or a desperate kamikaze than employing equally
aggressive response.

\begin{table}
\protect\caption{\label{tab:NCG-Kamikaze-matrix-form-1}The typical setup of the \emph{Kamikaze}
game with two players. Player-1 is the ``kamikaze'' and Player-2
is the defender.}

\centering{}%
\begin{tabular}{ccrccc}
\noalign{\vskip0.1cm}
\multirow{4}{0.1cm}{} & \multirow{2}{1.5cm}{\emph{Kamikaze}} &  & \multicolumn{2}{c}{Player-2} & \multirow{4}{0.7cm}{}\tabularnewline
 &  &  & \emph{C} & \emph{D} & \tabularnewline
\cline{3-5} 
\noalign{\vskip0.1cm}
 & \multirow{2}{1.5cm}{Player-1} & \emph{C} & (2,3) & (1,4{*}) & \tabularnewline
 &  & \emph{D} & (4{*},\emph{1}) & (\textbf{3}{*},\textbf{\emph{2}}{*}) & \tabularnewline
\cline{3-5} 
\end{tabular}
\end{table}

The concepts described along the strategic analysis and ``rationalization''
of the players in games like \emph{Hostage Situation} and \emph{Kamikaze}
illustrate how a seemingly irrational course of actions can be easily
explained and even classified as rational behavior, if the proper
utility functions are employed. In other words, if the utility of
each and every player is defined correctly, then all players in any
game can be labeled as ``rational'' ones. This proposition is often
referred to as ``rational irrationality'' (valid/explainable behavior),
rather than ``irrational rationality'' (incomprehensible behavior)
\cite{Mero1998}.

\parbox[c]{0.95\columnwidth}{%
\begin{shaded}%
\emph{Summary:}
\begin{itemize}
\item \emph{Utility} is the generalized cost/gain function of a player in
a specific game, depending on the outcomes but including any ``hidden''
regards and side-effects.
\item Given a specific \emph{utility function}, a player's \emph{incentives}
emerge naturally as the rational behavior of the underlying payoff-optimization
process.
\item A player's behavior may seem ``irrational'' if its utility function
is incomplete; given a properly defined utility function, a player's
behavior can always be labeled as rational per se.
\item \emph{Hostage Situation} and \emph{Kamikaze} are two examples of (asymmetric)
stand-off games where the notion of ``rational irrationality'' is
fully explained via proper definition of the corresponding utility
functions for the assaulter.\end{itemize}
\end{shaded}%
}

\section{The frontier}

This paper included only some of the most basic concepts of Game Theory,
including solution methods and representations of typical games of
special interest, like \emph{Chicken} and \emph{Prisoner's Dilemma}.
However, these are only a scratch on the surface of what lies beneath,
the rigorous mathematical theory and the complex, some still unsolved,
problems in this extremely interesting and useful scientific area.

All the games and setups presented thus far was somewhat ``too perfect'',
too simple compared to real-world situations of conflict. There are
few cases where only two players are involved, their moves are full
observable and their incentives clear and consistent. In most conflicts,
groups of players are spiraling in alternating rounds competing and
cooperating, each knowing its own utility function and very little
about the others', while signaling, third-party credibility assertions
and continuous bargaining are common things. Is there really a way
Game Theory can address all these aspects in the same clarity, mathematical
robustness and universality as is does with simple cases of zero-sum
and nonzero-sum games like the ones presented previously?

The short answer is ``No''. Game Theory is the mathematical way
to approach some of the most complex problems the human mind has ever
encountered. For example, what are the prerequisites, the dynamics
and the survivability of the evolution of cooperation as a strategy,
in human or animal societies? What is the asymptotic behavior of such
``cooperative'' groups? Can they survive in an environment of pure
competition? These issues are addressed in other aspects of the theory,
namely the \emph{Evolutionary Stable Strategies} (ESS), not analyzed
in this study. In short, ESS are patterns of behavior in games of
pure competition and/or possible cooperation, such as the \emph{Prisoner's
Dilemma}, that not only may emerge spontaneously but also survive
as optimal strategies in iterative games. \emph{Tit-for-Tat} \cite{Axelrod1984,Axelrod1988}
is such an example of ESS in iterated \emph{Prisoner's Dilemma}: cooperation
can emerge spontaneously given a set of conditions, primarily (a)
players ``start nicely'', (b) continue with reciprocity, (c) don't
know when the game finishes. Although it seems simple enough, spontaneous
cooperation in conflict situations is one of the most intriguing and
theoretically complex problems in Game Theory today.

In a slightly simpler scenario, a player may be involved in a game
with another player, while \emph{at the same time} its strategic choices
are relevant to a second game, with some other player. For example,
a politician may be in a ``bargain'' with voters, trying to gain
their support by promising specific actions if elected, while at the
same time a second ``bargain'' may be taking place in parallel with
the party's main policies and governmental plan if it comes to power.
If some of that politician's promises are on conflict with the party's
main lines, then as a player is involved in what is called a \emph{two-level
game} \cite{Putnam1978tlg,Putnam1988tlg}. This form of gaming was
first proposed by Putnam in the late '70s and it models two-level
or multi-level conflict situations in general, where the strategic
choices of a player affect two or more simultaneous games. The solution
concepts and equilibria are not much different than those of simple
games, but now a strategy is optimal and produces a stable outcome
only if it is such \emph{simultaneously} in all these games.

Another very interesting aspect of gaming in general is the evolution
of strategies and each player's behavior as each observes the others'
moves. In single-step games, the Minimax solution (pure or mixed)
is the one that dictates the optimal strategy for each player. The
concept of iterative gaming is much more general, since it includes
cases where the same players may face one another in the same single-step
games many times in the future. In this case, Nash equilibria predict
the most probable outcomes with much better accuracy. But the knowledge
that there will be a ``next round'', especially when players alternate
moves and one can observe the other before making its own (e.g. in
Chess), then the game analysis can expand to two or more steps ahead.
In practice, the player does not only take into account the strategic
choices available to the opponent(s) but also the ``what if'' combinations
of moves-countermoves. Hence, the corresponding game matrix includes
these combinations of composite states on the opponent(s) side and
the payoffs are estimated accordingly. This type of composite multi-step
setup is often referred to as a \emph{metagame} \cite{Thomas1984}.
The extended-form representation of metagames is more natural than
the analytical (matrix) form, but the identification of equilibria
and solutions is somewhat less straight-forward.

Some games involve elements of chance regarding the game's state or
partial information regarding the observability of each player's moves.
In such games of imperfect information, modeling via a game matrix
or a tree-graph can be problematic, since many of the paths may be
mutually exclusive and not just alternative choices. In the '60s,
very early on in the history of Game Theory, Harsanyi introduced the
so-called \emph{Harsanyi transformation} \cite{Harsanyi1962tr,Harsanyi1967tr,Montet2003}
for transforming a game of incomplete information to an equivalent
game of complete but imperfect information. This may not seem much,
but in reality there is a very distinct and important difference between
them. If a random event dictates the exact structure and payoffs of
the games, perhaps even the strategic behavior of the players, then
the analysis of such a game is inherently a very difficult task. On
the other hand, the Harsanyi transformation models this random event
as a deterministic one, removing the element of chance and introducing
the notion of ``hidden'' information about it. In practice, this
results in creating multiple variations of the game, one for each
possible configuration, and treating them separately. After they are
individually analyzed, solutions and equilibria are combined together
within a probabilistic framework, introducing the more generalized
concept of Bayesian Nash equilibria \cite{Montet2003}.

In real-world conflict situations it is not uncommon that one or some
of the players have a different knowledge or ``view'' of the game
structure, its payoffs and the other players' preferences. This means
that each player acts upon its own payoff matrix, possibly very different
in structure and values than the one used by the other players. Of
course, all players are involved in the same, single game and the
payoffs on each outcome is effectively a single one, despite each
player's unique view of the game. This is extremely important if some
of the players have a more complete view of the game, i.e., when they
address the game as one of (almost) complete information, while some
opponents address it as one of incomplete information. These special
types of conflict are often referred to as \emph{hypergames} \cite{Vane2006hg,Bennett1979hg}.
Introduced by Bennett and Dando in late '70s and later revised in
the '00s by Vane and others, hypergames is a very efficient way to
describe games of asymmetric information between players by employing
different variations of the game matrix or tree-graph, according to
each player's view. In practice, hypergames are treated the same way
as simple games, with each player deciding its strategic choices according
to its own view and, subsequently, combining the (partial) outcomes
together.

Game Theory is a vast scientific and research area, based almost entirely
on Mathematics and some experimental methods, with applications that
vary from simple board games and auctions to Evolutionary Psychology
and Sociology-Biology in group behavior of humans and animals. Although
real-world situations reveal that sometimes its predictive value is
limited, the robust theoretical framework and solution concepts provide
an extremely valuable set of tools that clarifies the inner workings
and dynamics of conflict situations.

\parbox[c]{0.95\columnwidth}{%
\begin{shaded}%
\emph{Summary:}
\begin{itemize}
\item In accordance to Nash's bargaining theorem, cooperation can \emph{emerge
spontaneously}, even in competitive games, when a specific set of
pre-requisites are satisfied. 
\item \emph{Evolutionary stable strategies} (ESS) are patterns of behavior
in games of pure competition and/or possible cooperation that survive
as optimal strategies in iterative games.
\item In \emph{two-level games}, a player may be involved in a game with
another player, while \emph{at the same time} its strategic choices
are relevant to a second game, with some other player. 
\item \emph{Metagames} are multi-step game setups where the corresponding
game matrix includes combinations of ``what if'' composite states,
regarding the future strategic choices of the opponent(s).
\item The \emph{Harsanyi transformation} is used in games of incomplete
information, e.g. when the game structure and payoffs depend on some
random event, to transform it to an equivalent game of complete but
imperfect information.
\item \emph{Hypergames} is a very efficient way to describe games of asymmetric
information between players by employing different variations of the
game matrix or tree-graph, according to each player's view.
\item In general, Game Theory is a vast scientific and research area with
robust theoretical foundation that can be used as a predictive tool,
as well as (mostly) an extremely valuable approach to analyze conflict
situations.\end{itemize}
\end{shaded}%
}

\textbf{\textit{\vspace{12mm}
}}

\textbf{\emph{Acknowledgement:}}\emph{ This work is dedicated to John
F. Nash, pioneer and mathematical genius, who was killed earlier this
month on May 23th 2015 in a car accident along with his wife Alicia.
His inspirational work and breakthrough ideas has changed Game Theory
and Economics forever.}

\textbf{\textit{\newpage{}}}

\bibliographystyle{plain}
\bibliography{game-theory-misc}

\textbf{\textit{\newpage{}}}\textbf{\small{}Harris Georgiou}{\small{}
received his B.Sc. degree in Informatics from University of Ioannina,
Greece, in 1997, and his M.Sc. degree in Digital Signal Processing
\& Computer Systems and Ph.D. degree in Machine Learning \& Medical
Imaging, from National \& Kapodistrian University of Athens, Greece,
in 2000 and 2009, respectively. Since 1998, he has been working with
the Signal \& Image Processing Lab (SIPL) in the Department of Informatics
\& Telecommunications (DIT) at National \& Kapodistrian University
of Athens (NKUA/UoA), Greece, in various academic and research projects.
In 2013-2015 he worked as a post-doctorate associate researcher with
SIPL in sparse models for distributed analysis of functional MRI (fMRI)
signals. He has been actively involved in several national and EU-funded
research \& development projects, focusing on new and emerging technologies
in Biomedicine and applications. He has also worked in the private
sector as a consultant in Software Engineering and Quality Assurance
(SQA, EDP/IT), as well as a faculty professor in private institutions
in various ICT-related subjects, for more than 17 years. His main
research interests include Machine Learning, Pattern Recognition,
Signal Processing, Medical Imaging, Soft Computing, Artificial Intelligence
and Game Theory. He has published more than 65 papers and articles
(43 peer-reviewed) in various academic journals \& conferences, open-access
publications and scientific magazines, as well as two books in Biomedical
Engineering \& Computer-Aided Diagnosis and several contributions
in seminal academic textbooks in Machine Learning \& Pattern Recognition.
He is a member of the IEEE and the ACM organizations and he has given
several technical presentations in various countries.}{\small \par}
\end{document}